\documentclass[lettersize,journal]{IEEEtran}
\IEEEoverridecommandlockouts

\usepackage{cite}
\usepackage{amsmath,amssymb,amsfonts}
\usepackage{algorithmic}
\usepackage{graphicx}
\usepackage{amsfonts}
\usepackage{amssymb}
\usepackage{textcomp}
\usepackage[T1]{fontenc}
\usepackage{longtable} 
\usepackage{supertabular} 
\usepackage{xcolor}
\usepackage{subcaption}
\usepackage{tabularx}
\usepackage{comment}
\usepackage{booktabs}
\usepackage{mathrsfs} 
\usepackage{float} 
\usepackage{booktabs}

\usepackage{algorithmic}\makeatletter
\usepackage{algorithm}  
\usepackage{multirow}
\usepackage{makecell}
\usepackage{hyperref}

\allowdisplaybreaks
\def\BibTeX{{\rm B\kern-.05em{\sc i\kern-.025em b}\kern-.08em
    T\kern-.1667em\lower.7ex\hbox{E}\kern-.125emX}}
\begin{document}

\title{Enhancing High-Speed Cruising Performance of Autonomous Vehicles through Integrated Deep Reinforcement Learning Framework
 \\

\author{Jinhao Liang, \textit{Member, IEEE}, Kaidi Yang, Chaopeng Tan, Jinxiang Wang, \textit{Member, IEEE}, Guodong Yin, \textit{Senior Member, IEEE}
\thanks{This research was supported by the Singapore Ministry of Education (MOE) under NUS Start-Up Grant (A-8000404-01-00). This article
solely reflects the opinions and conclusions of its authors and not the Singapore MOE or any other entity. (Corresponding author: Kaidi Yang).

Jinhao Liang, Kaidi Yang, and Chaopeng Tan are with the Department of Civil and Environmental Engineering, National University of Singapore, Singapore, 119077, (E-mail: \{jh.liang, kaidi.yang, tancp\}@nus.edu.sg).

Jinxiang Wang and Guodong Yin are with the School of Mechanical Engineering, Southeast University, Nanjing, 211189, China. (E-mail: \{wangjx,ygd\}@seu.edu.cn).
}
\thanks{This work has been submitted to the IEEE for possible publication. Copyright may be transferred without notice, after which this version may no longer be accessible.}

}}
\maketitle

\begin{abstract}
High-speed cruising scenarios with mixed traffic greatly challenge the road safety of autonomous vehicles (AVs). Unlike existing works that only look at fundamental modules in isolation, this work enhances AV safety in mixed-traffic high-speed cruising scenarios by proposing an integrated framework that synthesizes three fundamental modules, i.e., behavioral decision-making, path-planning, and motion-control modules. Considering that the integrated framework would increase the system complexity, a bootstrapped deep Q-Network (DQN) is employed to enhance the deep exploration of the reinforcement learning method and achieve adaptive decision making of AVs. Moreover, to make AV behavior understandable by surrounding HDVs to prevent unexpected operations caused by misinterpretations, we derive an inverse reinforcement learning (IRL) approach to learn the reward function of skilled drivers for the path planning of lane-changing maneuvers. Such a design enables AVs to achieve a human-like tradeoff between multi-performance requirements. Simulations demonstrate that the proposed integrated framework can guide AVs to take safe actions while guaranteeing high-speed cruising performance.

\end{abstract}


\section{Introduction}
\IEEEPARstart{A}{utonomous} vehicles (AVs) have been widely recognized as a promising technology with enormous potential for improving traffic efficiency, road safety, and energy consumption in intelligent transportation systems \cite{zhao2020enhanced, phan2020intelligent}. Equipped with advanced sensors, such as cameras and LiDARs, AVs demonstrate the capability of perceiving the environment and serving as control actuators for improving traffic flow \cite{van2018autonomous}. However, the penetration rates of AVs can only gradually increase as technological maturity and public acceptance improve, and hence AVs are expected to operate in mixed traffic scenarios together with human-driven vehicles (HDVs). Within this context, the inherent unpredictability of HDVs emphasizes the critical need to investigate the safe control of AVs in mixed-traffic environments. 
On the one hand, this requires AVs to systematically consider their autonomous driving capabilities and the behavior of surrounding vehicles. On the other hand, it is crucial to make the decisions of AVs understandable by surrounding HDVs to prevent unexpected operations caused by misinterpretations.
 
Most studies divide the autonomous driving task into three sequentially executed modules, i.e., the behavioral decision-making, path-planning, and motion-control modules, and seek to enhance the safety of AVs by improving each of them in isolation 
\cite{ye2019automated, wang2023adaptive, hu2020fuzzy}. The behavioral decision-making module, taking into account road rules and traffic conditions, is first executed to help AVs choose between lane keeping and lane changing, based on methods such as finite state machines \cite{hu2022decision}, behavior trees \cite{giunchiglia2019conditional}, and deep reinforcement learning (DRL) \cite{zhu2018human, balal2016binary, xiong2018decision}. It is worth noting that DRL-based methods have recently attracted increasing research attention for the behavioral decision-making module, thanks to its ability to make robust decisions efficiently in complex traffic conditions. Then, if lane changing is chosen, the path-planning module is executed to generate a collision-free trajectory, which is 
 typically achieved by formulating an optimization problem \cite{li2021optimization, han2017multi, claussmann2019review, scheffe2022sequential, fayazi2018mixed}. 
Finally, the motion-control module allows AVs to implement behavioral decisions and maintain precise and stable movement along planned trajectories. Some model-based algorithms, such as linear quadratic regulator \cite{xu2019design}, sliding mode control \cite{hwang2017path}, and robust H-$\infty$ control \cite{chen2019human} are widely used to guarantee autonomous tracking performance. Recently, Model Predictive Control (MPC) \cite{wang2020autonomous, liang2021distributed, ji2016path} has received considerable attention due to its ability to explicitly handle system constraints. 

However, the above-mentioned studies enhance AV safety only focusing on individual modules in isolation without considering the coupling between these modules.
Intuitively, without accounting for the lower-level modules, the upper-level decisions may not be optimal or even physically feasible \cite{paden2016survey}. To address this issue, several studies have been conducted to achieve integrated control by combining some of these modules. In \cite{peng2022integrated}, a hierarchical reinforcement learning framework is proposed for the integrated control of lane-changing decisions and velocity planning. In \cite{naveed2021trajectory}, an enhanced double DQN algorithm is employed to generate binary actions to choose between lane keeping and lane changing. Reference \cite{li2022combining} extends \cite{naveed2021trajectory} by further introducing a path-planner to provide a safety guarantee in generating lane-changing maneuvers. 
Nevertheless, to the best of our knowledge, the motion-control module is rarely considered in these integrated frameworks or simply regarded as a kinematic model   \cite{peng2022integrated, naveed2021trajectory, li2022combining}. 
Ignoring the motion-control module can render the planned path or behavioral decision infeasible, which is especially possible in high-speed cruising scenarios where the physical constraints of the vehicles are more likely active, making the kinematic models inaccurate and 
leading to a deterioration in motion-control performance and safety concerns. Therefore, it is crucial to incorporate a motion-control module into the integrated control framework \cite{hawke2020urban}. 

Moreover, even if safety can be enhanced from a technological perspective, human drivers may not understand the unexpected decisions made by AVs. 
This lack of understanding could potentially lead to undesirable HDV operations and increase the risk of collisions~\cite{liu2019evaluating}. 
Therefore, enhancing the safety of autonomous driving also requires ensuring that HDVs understand the behavior of AVs, typically characterized by their trajectories.  
One promising solution to improve the interpretability of autonomous driving is to make AV behavior human-like  \cite{ruijten2018enhancing,liu2019evaluating}. 
Current research on achieving human-like behavior in AVs aims to emulate human drivers' characteristics during navigation, thereby enhancing AVs' adaptability in complex environments \cite{wei2021human}. This pursuit seeks to foster greater compatibility and trust between AVs and HDVs on the road. 

The commonly employed approach to achieve human-like capabilities involves the utilization of machine learning models trained on extensive datasets to simulate human behavior. In \cite{emuna2020deep}, a model-free DRL framework was proposed to learn obstacle-avoidance behavior from skilled drivers. This framework encompassed a data-driven approach to integrate the expertise of human drivers and a rule-based method to encode fundamental driving rules. Reference \cite{xu2020learning} proposed heuristic and learning methods for training human-like lane-changing maneuvers based on naturalistic driving data. However, traditional machine learning methods suffer from several drawbacks, including relying solely on trial-and-error exploration and having limited generalization in new driving scenarios. Inverse Reinforcement Learning (IRL) offers advantages, such as the utilization of expert behavior demonstrations to learn task-relevant features and better adaptation to dynamic environments \cite{naumann2020analyzing}. These advantages enable AVs to better understand and mimic human-like behaviors through demonstrations rather than rely solely on labeled data. Furthermore, unlike the conventional IRL method \cite{kebria2019deep} that focuses on directly learning human driving actions,  we learn the reward functions of human drivers' lane-changing maneuvers in this study to understand the tradeoff between various control objectives. This design helps incorporate the IRL method into the path-planning process to generate a human-like path.

\emph{Statement of Contribution}. The main contributions of this paper are two-fold. First, we enhance AV safety by proposing an integrated AV control framework in the high-speed cruising scenario that synthesizes the three fundamental modules of autonomous driving, i.e., behavioral decision-making, path-planning, and motion-control modules. Specifically, the behavioral decision-making module leverages Bootstrapped DQN to determine whether to perform lane-keeping or lane-changing, considering the interactions with the path-planning and motion-control modules. The Bootstrapped DQN can effectively enhance the deep exploration ability, which is important with the added complexity brought by the integration with two lower-level modules. 
Second, we 
devise an IRL-based approach to learn skilled drivers' reward function for planning lane-changing paths, which 
characterizes the human-like tradeoff between 
various considerations (e.g., vehicle safety, driving comfort, and travel efficiency).

The rest of this paper is organized as follows. Section II introduces the overall framework, while Sections III, IV, and V present the bootstrapped DQN-based decision-making module, the IRL-inspired path-planning module, and the MPC-based motion-control module, respectively. Section VI analyzes the simulation results to highlight the significance of the proposed integrated framework. Section VII concludes the paper.

\section{Problem Formulation and Overall Framework} 
\subsection{Problem Formulation}
\begin{figure}[ht]
	\centering
	\includegraphics[width=2.4in]{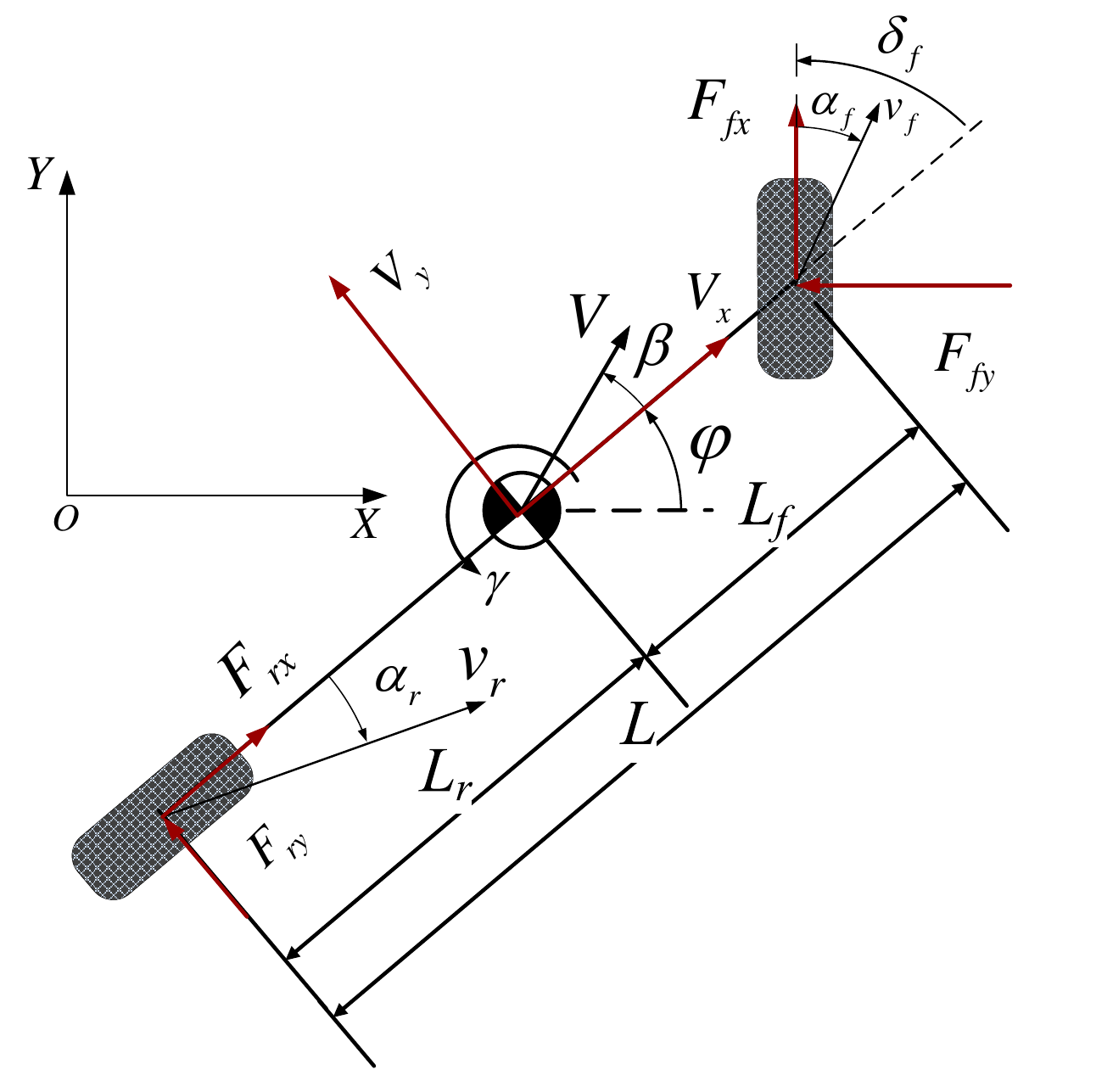}
	\caption{Vehicle dynamics model.}\label{fig1:vehicle dynamics model}
\end{figure}

Let us consider a highway scenario where an AV seeks to optimize its high-speed cruising performance. We assume that all surrounding vehicles around this AV are HDVs, which is typical in the early deployment stages of AVs with very low penetration rates. The driving behavior of these surrounding HDVs is characterized by the Intelligent Driver Model (IDM) for car following and the MOBIL model for lane changing, both of which are widely adopted in transportation research \cite{peng2022integrated}. The vehicle dynamics of AVs are modeled using a bicycle model (see Fig. \ref{fig1:vehicle dynamics model}) with configuration parameters obtained from a high-fidelity vehicle model in the commercial software Carsim, as detailed in Table \ref{tab1:vehicle model parameters}. 

This study aims to devise a controller for the longitudinal (i.e., lane-keeping) and lateral (i.e., lane-changing) movements of the AV in high-speed cruising scenarios. Hence, the AV's control input is defined by $v=\left[a_{x}, \delta_{f}\right]^{T}$, where $a_{x}$ and $\delta_{f}$  are the longitudinal acceleration and front-wheel steering input, respectively. The AV's state is defined by $\zeta=\left[V_{x}, V_{y}, \varphi, \gamma, X, Y\right]^{T}$, where $V_{x},~V_{y},~\varphi,~\gamma,~X$, and $Y$ represent the longitudinal velocity, lateral velocity, yaw angle, yaw rate, global longitudinal position, and global lateral position, respectively. 

With these notations, the vehicle's position under the global coordinate system can be modeled as
\begin{subequations}  
 \label{eq3:Vehicle global positions}
  \begin{align}
\dot{X}&=-V_{y} \varphi+V_{x} 
\\
\dot{Y}&=V_{x} \varphi+V_{y}
  \end{align}
\end{subequations}%

When the vehicle operates with a small steering input, the dynamics of the longitudinal velocity, lateral velocity, and yaw angle can be represented as follows \cite{liang2022mas}.
\begin{subequations} \label{eq1:vehicle dynamics model}
    \begin{align} 
M\left(-V_{y} \gamma+\dot{V}_{x}\right)&= F_{f x} + F_{r x}, \\
M\left(V_{x} \gamma+\dot{V}_{y}\right)&=F_{f y} + F_{r y}, \\
J_{z} \dot{\gamma}&=-L_{r} F_{r y}+L_{f} F_{f y},
\end{align}
\end{subequations}
where $M$ represents the vehicle mass, $J_z$ represents the inertia yaw moment of the vehicle, $F_{f y}$ and $F_{f x}$ represent the lateral and longitudinal forces of the front tire, respectively,  $F_{r y}$ and $F_{r x}$ represent the lateral and longitudinal forces of the rear tire, respectively, and $L_{f}$ and $L_{r}$ represent the distances from the center of gravity (CG) to the front axle and rear axle, respectively. As shown in Fig.~\ref{fig1:vehicle dynamics model}, the relation between these variables can be described by  
\begin{equation} \label{eq3:veh_dy}
  \begin{split}
& F_{f x}+F_{r x}=M a_{x}, F_{f y}=2 K_{f} \alpha_{f}, F_{r y}=2 K_{r} \alpha_{r} \\
& \alpha_{f}=\frac{L_{f} \gamma+V_{y}}{V_{x}}-\delta_{f}, \alpha_{r}=\frac{-L_{r} \gamma+V_{y}}{V_{x}}
  \end{split}
\end{equation}
where $K_{f}$ and $K_{r}$ represent the cornering stiffness of front and rear tires, respectively, $\alpha_{f}$ and $\alpha_{r}$ represent the slip angles of front and rear tires, respectively.  $\beta$ is the vehicle sideslip angle and calculated by $V_{y} / V_{x}$. 

By combining Eq.(\ref{eq3:Vehicle global positions})-Eq.(\ref{eq3:veh_dy}), the vehicle dynamics can be written as follows.
\begin{align} \label{eq4:state-space equation}
\dot{\zeta}=A \zeta+B v
\end{align}
where
\begin{align*}
A\!&=\!\begin{bmatrix}
    0 & 0 & 0 & V_{y} & 0 & 0 \\ 0 & \frac{2\left(K_{r}+K_{f}\right)}{M V_{x}} & 0 & \frac{2\left(K_{f} L_{f}-K_{r} L_{r}\right)}{M V_{x}}-V_{x} & 0 & 0 \\ 0 & 0 & 0 & 1 & 0 & 0 \\ 0 & \frac{2\left(C_{f} l_{f}-C_{r} l_{r}\right)}{J_{z} V_{x}} & 0 & \frac{2\left(K_{r} L_{r}^{2}+K_{f} L_{f}^{2}\right)}{J_{z} V_{x}} & 0 & 0 \\ 1 & 0 & -V_{y} & 0 & 0 & 0 \\ 0 & 1 & V_{x} & 0 & 0 & 0\end{bmatrix}, \\
B\!&=\!\left[\begin{array}{cccccc}1 & 0 & 0 & 0 & 0 & 0 \\ 0 & \frac{-2 K_{f}}{M} & 0 & \frac{-2 K_{f} L_{f}}{J_{z}} & 0 & 0\end{array}\right]^{T}.
\end{align*}

\begin{table}[ht]
	\begin{center}
		\caption{Vehicle Configuration Parameters}
		\begin{tabular}{ c  c  c}
			\hline
			\toprule  
			Symbol & Description & Value[units] \\
			\midrule  
			$M$ & Vehicle mass & $1274~(\mathrm{kg})$ \\
$J_{z}$ & Inertia yaw moment of the vehicle & $606.1~\left(\mathrm{kg} \cdot \mathrm{m}^{2}\right)$ \\
$L_{f}$ & Distance from front axle to CG & $1.016~(\mathrm{m})$ \\
$L_{r}$ & Distance from rear axle to CG & $1.562~(\mathrm{m})$ \\
$K_{f}$ & Cornering stiffness of front tire & $85000~(\mathrm{N} / \mathrm{rad})$ \\
$K_{r}$ & Cornering stiffness of rear tire & $112000(\mathrm{~N} / \mathrm{rad})$ \\
			\bottomrule 
			\hline 
		\end{tabular}
  \label{tab1:vehicle model parameters}
	\end{center}
\end{table}
 
\begin{figure*}[t]
	\centering
	\includegraphics[width=6.2 in]{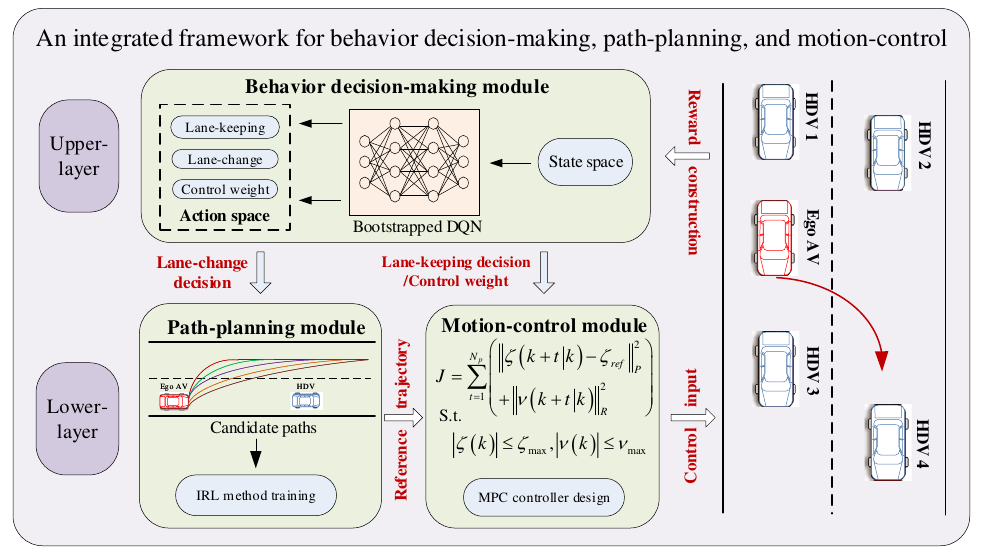}
	\caption{The Systematic framework of the proposed method.} \label{fig2:The Systematic framework of the proposed method}
\end{figure*}
 
\subsection{Overall Framework}
This section presents an integrated framework (see Fig. \ref{fig2:The Systematic framework of the proposed method}) to systematically combine the decision-making, path-planning, and motion-control modules. The decision-making module chooses between lane-keeping and lane-changing maneuvers, using a DRL agent that takes into account high-speed cruising performance and collision avoidance. If lane keeping is chosen, the DRL agent generates an acceleration/deceleration control signal.
If lane changing is chosen, the path-planning module generates an efficient and human-like path that effectively trade off vehicle safety, driving comfort, and travel efficiency, with the corresponding weights learned from experienced drivers using an IRL algorithm. 
Finally, the motion-control module calculates the control signals with a model predictive controller that tracks the planned path while ensuring the physical feasibility by fully considering the vehicle dynamics constraints and actuator saturation.

Note that, different from existing works that combine only the decision-making module with the path-planning module, we further incorporate the motion-control module into the integrated framework. Such an integration involves two aspects. First, the DRL agent in the behavioral decision-making module generates not only the behavioral decisions but also the control weights that will be used in the motion-control module. Such a treatment systematically aligns the control objectives of the motion-control module with the other two modules. Second, the motion-control module is incorporated into the DRL agent's training process, thereby ensuring that the trained DRL agent can generate physically feasible actions. 

\section{Bootstrapped DQN-based Behavioral Decision-Making for the Autonomous Vehicle} \label{Bootstrapped DQN-based Decision-Making for the Autonomous Vehicle}
This section presents a bootstrapped DQN method for the behavioral decision-making module in the integrated framework. Here, we leverage the bootstrapped DQN method to enhance the deep exploration ability to address the added complexity of incorporating the other two modules into the training process. 
\subsection{MDP Formulation} \label{sec_MDP}
The decision-making process can be described as a Markov Decision Process (MDP), denoted by a tuple $\mathcal{M}=(\mathcal{S}, \mathcal{A}, P, R)$, where $\mathcal{S}$, $\mathcal{A}$, $P$, and $R$ are the state space, action space, state transition probability, and reward function, respectively. The details are presented below.

\vspace{0.5em}
\noindent \emph{1) State Space}. The state space comprises the vehicle states of the ego AV and its relative vehicle states to the HDVs. The state of the MDP at decision step $i$ is represented by $S_i=\left(\vartheta_{ev,i}, \{\vartheta_{ev,i} - \vartheta_{m,i}\}_{m=1}^n\right)$, where $\vartheta_{ev,i}=[Y_{ev,i}, X_{ev,i},V_{ev,i}]^T$ denotes the states (i.e., lateral position, longitudinal position, and velocity) of the ego AV at decision step $i$, and $\vartheta_{m,i}=[Y_{m,i}, X_{m,i},V_{m,i}]^T$ denotes the corresponding states of the surrounding HDV $m$ at decision step $i$.

\vspace{0.5em}
\noindent \emph{2) Action Space}. 
 In a high-speed cruising scenario, the DRL agent in the behavioral decision-making module makes a choice between lane-keeping and lane-changing maneuvers while simultaneously generating the control weights of the motion-control module associated with these driving maneuvers. Therefore, the action for an ego AV comprises lane-keeping and lane-changing decisions, as well as control weights for the motion-control module. Specifically, as illustrated in Table \ref{tab2:Action space}, the action for an ego AV can be represented as an 8-dimensional vector $u = [u_{\rm{acc}}, u_{\rm{mod}}, u_{\rm{dcc}}, u_{\rm{lc}}, P_1,R_1,P_2,R_2]$ with $[u_{\rm{acc}},u_{\rm{mod}},u_{\rm{dcc}},u_{\rm{lc}}]$ being a one-hot vector (i.e., only one element being one and the others being zeros).  Here, lane-keeping decisions $u_{\rm{acc}}$, $u_{\rm{mod}}$, and $u_{\rm{dcc}}$ are binary variables representing the selection of acceleration with a rate of $2\,\mathrm{~m} / \mathrm{s}^{2}$, no acceleration, and deceleration with a rate of $-2\,\mathrm{~m} / \mathrm{s}^{2}$, respectively, where the value of $2\,\mathrm{~m} / \mathrm{s}^{2}$ is chosen to consider vehicle dynamics and road adhesion limits. The lane-changing decision variable $u_{\rm{lc}}$ is also a binary variable with $u_{\rm{lc}}=1$ indicating the selection of lane-changing maneuvers and the activation of the path-planning module. The weights $P_{1}$, $R_{1}$, $P_{2}$ and $R_{2}$ represent the coefficients of various criteria in the objective function of the motion-planning module, the specific forms of which are defined in Section IV-C.

\begin{table}[ht] 
	\begin{center}
		\caption{AV’s Action Space} 
		\begin{tabular}{ c c c c c c}
			\hline
           \multirow{2}{0.5cm}{\centering Action space}  &  \multicolumn{3}{c}{Lane-keeping decision}   & \multirow{2}{1.7cm}{\centering {Lane-changing decision}}  &\multirow{2}{0.5cm}{\centering {Control weights}}  \\
           \cline{2-4}
           ~ & \multicolumn{1}{p{1.1cm}}{\centering Acceleration} & \multicolumn{1}{p{1.1cm}}{\centering Moderate} &\multicolumn{1}{p{1.1cm}}{\centering Deceleration} &~ & ~ \\
             \hline
             \multirow{2}{0.5cm}{\centering Value} & $+2$ & $0$ & $-2$ & \multirow{1}{1.7cm}{\centering {Path-planning process}}  &\multirow{1}{0.5cm}{\centering {$P_1, P_2$}} \\
             ~ & $(m/s^2)$ & $(m/s^2)$ & $(m/s^2)$ &~ & \multirow{1}{0.5cm}{\centering {$R_1, R_2$}} \\
             \hline
		\end{tabular}\centering \label{tab2:Action space}
	\end{center}
\end{table}

\vspace{0.5em}
\noindent \emph{3) Reward}.
This work incorporates high-speed driving performance and collision avoidance guarantees into the construction of the reward function, defined as follows: 
\begin{align} \label{eq5:Reward}
r=\omega_{sped} r_{sped}+\omega_{col} r_{col}+\omega_{lanc} r_{lanc},
\end{align}
where $r_{sped}$, $r_{col}$, and $r_{lanec}$ represent the high-speed incentive, collision penalty, and frequent lane-changing penalty, respectively, with the forms specified as Eq.(\ref{eq6:High-speed cruising Reward})-Eq.(\ref{eq8:Reward:discourage frequent lane-changing}), and $\omega_{sped}, \omega_{col}, \omega_{lanc}$ represent the corresponding coefficients that are chosen as $20,-5$, and $-0.1$, respectively, after experiments. 

\begin{align} 
r_{sped}&=\frac{V_{e v, i}-V_{\min }}{V_{\max }-V_{\min }} ,\label{eq6:High-speed cruising Reward} \\
r_{col}&=\sum_{m=1}^{n} \exp \left[\sigma_{1}\left(Y_{e v,i}-Y_{m,i}\right)^{2}+\sigma_{2}\left(X_{e v,i}-X_{m,i}\right)^{2}\right],
\label{eq7:Collision-avoidance Reward} \\
r_{lanc}&=P_{\text {lane }} \cdot \exp \left(\frac{\left(Y_{e v,i}-Y_{\text {mid }}\right)^{2}}{2 \sigma_{3}^{2}}\right),
\label{eq8:Reward:discourage frequent lane-changing} 
\end{align}
 where Eq.(\ref{eq6:High-speed cruising Reward}) encourages the AV to drive faster, with $V_{e v, i}$, $V_{\min }$, and $V_{\max }$ indicating the ego AV velocity at decision step $i$, minimum velocity, and maximum velocity, respectively. The constraint $V_{e v, i} \in\left[V_{\min }, V_{\max }\right]$ is ensured in the lower motion-control module via the MPC controller. 
 Eq. (\ref{eq7:Collision-avoidance Reward}) employs an artificial potential field (APF) function to penalize the ego AV from being too close to surrounding HDVs to enhance its collision-avoidance ability in high-speed cruising scenarios, where $Y_{e v, i}$ and $X_{e v, i}$ represent the global lateral and longitudinal positions of ego AV at decision step $i$, while $Y_{m, i}$ and $X_{m, i}$ indicate the global lateral and longitudinal positions of HDV $m$ at decision step $i$. The scale factors $\sigma_{1}$ and $\sigma_{2}$ normalize the longitudinal and lateral spacing between vehicles. Eq.(\ref{eq8:Reward:discourage frequent lane-changing}) is a Gaussian-type reward with scale factor $\sigma_{3}$  implemented to discourage excessive lane-changing maneuvers to reduce the impact on congestion, where $Y_{\text {mid }}$ is the lateral coordinate of the lane centerline, and $P_{\text {lane }}$ normalizes the reward term to be between 0 and 1. During the training process, the scale factors are tuned to minimize collisions. After experiments, $\sigma_{1}$, $\sigma_{2}, \sigma_{3}$, and $P_{\text {lane }}$ are set by -3, -0.5, -0.1, and 2.5, respectively.  

\subsection{Bootstrapped DQN}
 The bootstrapped DQN is leveraged as the RL agent in this paper to enhance training efficiency, which provides exponential learning speed by approximating the population distribution using the distribution of bootstrapped samples $\Omega(\tilde{E})$, where the data set $\tilde{E}$ is obtained by sampling uniformly with replacement from the real data set $E$. 

\begin{figure}[ht]
	\centering
	\includegraphics[width=3in]{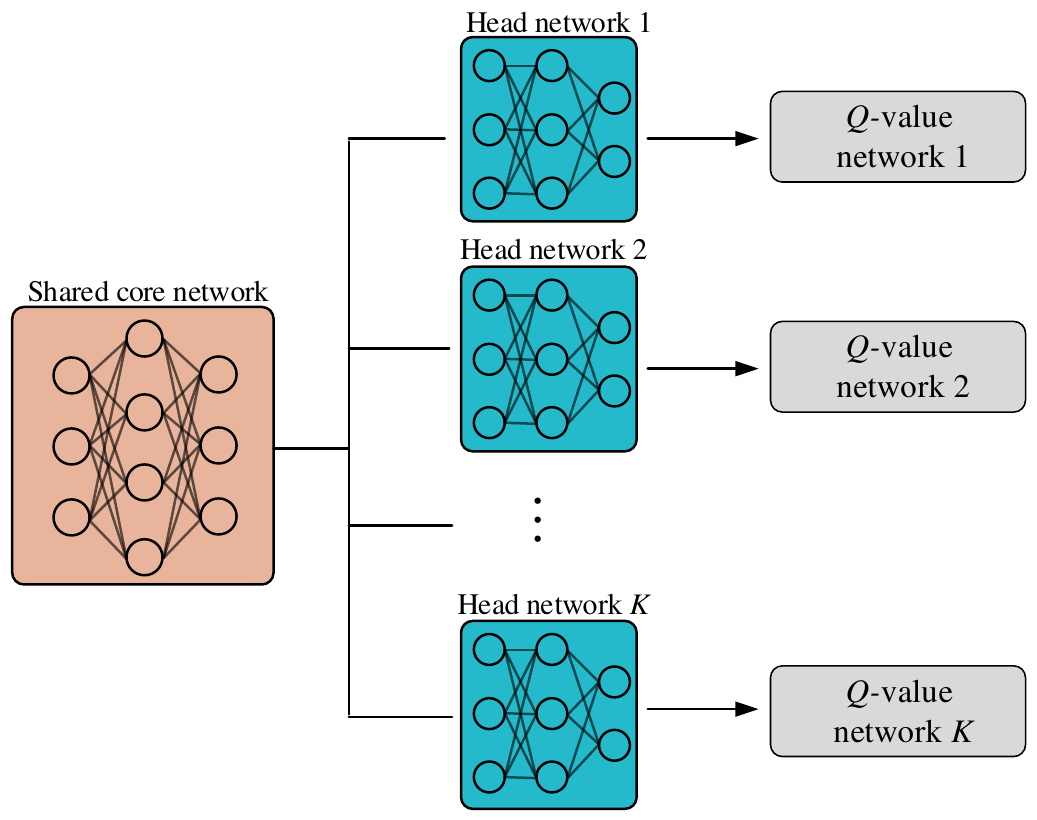}
	\caption{The network structure of the Bootstrapped DQN.}\label{fig3:The network structure of the Bootstrapped DQN}
\end{figure}

Fig. \ref{fig3:The network structure of the Bootstrapped DQN} illustrates the network structure of the bootstrapped DQN that consists of a shared core network and $k$ independent branches, each representing a head network. Each head network combined with the core network can be seen as a $Q$-value network, denoted by $Q^{k}$. Note that the training data for each head network only includes the bootstrapped sample with distribution $\Omega(\tilde{E})$. This allows $K$ independent $Q$-networks to collaborate and produce improved $Q$-value estimation results.

The bootstrapped DQN is implemented by generating a bootstrap mask to acquire a subsample each time after an action is selected. The bootstrap mask at decision step $i$, represented by a binary vector $d_i = \{d_i^k\}_{k=1}^K \in\{0,1\}^K$ with $K$ denoting the number of head networks, indicates which neural network to update with the current transition. Considering the bootstrap mask $d$, the state transition at decision step $i$ can be augmented by 
$\left(s_{i}, u_{i}, r_{i+1}, s_{i+1}, d_{i}\right)$ and stored in the experience buffer. Consequently, the gradient of the $k$-th $Q$-value network can be represented as follows.
\begin{align} \label{eq9:Gradient of the network}
g_{i}^{k}=d_{i}^{k}\left(y_{i}^{Q^{k}}-Q^{k}\left(s_{i}, u_{i} ; \theta_{i}^{k}\right)\right) \nabla_{\theta} Q^{k}\left(s_{i}, u_{i} ; \theta_{i}^{k}\right)
\end{align}
where
\begin{align}
y_{i}^{Q^{k}}=r_{i}+\gamma \max _{u} Q^{k}\left(s_{i+1}, u ; \theta^{-}\right)
\end{align}
where $\theta_{i}^{k}$ represents the $k$-th neural network parameters at decision step $i$, $y_{i}^{Q^{k}}$ represents the approximate target value at decision step $i$, and $ r_{i}$ and $\gamma$ refer to the interactive reward with the environment and the discount factor, respectively. $\max _{u} Q\left(s_{i+1}, u ; \theta^{-}\right)$ represents the action corresponding to the maximum action value of the target $Q$ network at the state $s_{i+1}$. During the training process, similar to the traditional Q-network \cite{li2022combining}, we introduce $\theta^{-}$ to denote the target network parameter, which satisfies the fixed $\theta^{-}=\theta_{i}^{k}$. It means the target network parameters $\theta^{-}$ are updated at a frequency of $\tau$ and remain fixed in between updates. Furthermore, gradient normalization is used to mitigate the impact of back-propagation from each $Q$-value network onto the shared core network. The normalized process is represented as follows.
\begin{align} \label{eq11:gradient normalization}
g_{i, n o r}^{k}=g_{i}^{k} / K
\end{align}

When an RL agent starts the training process and explores potential rewards from the environment, it first follows the basic DQN method. Note that a $Q$-value network is randomly chosen at each episode. After completing the training, the optimal policy can output a set of actions. Considering the diversity of the $Q$-value network in the Bootstrapped DQN structure, a voting mechanism is introduced to select the optimal action, whereby the action with the highest number of votes across $K$ different $Q$-value networks will be executed. The pseudo-code for the bootstrapped DQN is shown in Algorithm 1. 

\begin{algorithm}[ht]\label{tab1:The algorithm flow of the decision-making RL}
	\caption{The algorithm flow of the bootstrapped DQN}
	\renewcommand{\algorithmicrequire}{\textbf{Input:}}
	\renewcommand{\algorithmicensure}{\textbf{Initialize:}}
	\begin{algorithmic}[1]
		\REQUIRE Initialize $Q$-value networks $Q^{k} (k = 1...K$) and masking distribution $D$; target network update frequency $\tau$ and an empty experience buffer $B$.
		\ENSURE Network parameters $\theta_{0}^{k}$ and $\theta^{-}=\theta_{0}^{k}$.
		\FOR{each episode $T=1,2, \ldots, N$, where $N$ is the number of episodes,}
		\STATE Obtain state $s_{0}$ by interacting with the environment
        \STATE  Select a head network $Q^{k}$ to derive an action with $k$ chosen uniform at random from $\{1,2, \ldots, K\}$.
		\FOR{time step from $i=1,2 \ldots$ to the end of the episode,}
       \STATE Select an action $u_{i} \in \arg \max _{u} Q^{k}\left(s_{i}, u\right)$.   
        \STATE Transmit and execute action $u_i$ in path planning and motion control modules to obtain the next state $s_{i+1}$ and collect the interactive reward $r_{i}$ from the environment.
        \STATE Sample the mask $d_{i} \sim D$.
        \STATE Add the tuple $\left(s_{i}, u_{i}, r_{i+1}, s_{i+1}, d_{i}\right)$ to the experience buffer $B$.
        \IF{$i$ mod $\tau$=0,}
        \STATE $\theta^{-}=\theta_{i}^{k}$
        \ENDIF
        \IF{the experience buffer $B$ is full,}
        \STATE update $\theta_{i}^{k}$
		\ENDIF
      \ENDFOR
		\ENDFOR
	\end{algorithmic}
\end{algorithm}

\section{Path-planning Module Design via IRL Method}
After the DRL agent makes a lane-changing decision, the path-planning module is responsible for generating a human-like collision-free trajectory. The path-planning module includes two components: (i) a candidate path generation component that generates a set of possible paths and (ii) a selection component that chooses the optimal trajectory with a reward function learned from skilled human drivers via IRL to characterize human preferences over various control objectives. 

\subsection{Candidate Paths Generation}  
\vspace{0.5em}

\begin{figure}[ht]
	\centering
	\includegraphics[width=3in]{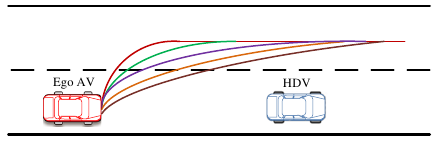}
	\caption{The candidate paths of path-planning scenario.} \label{Fig4:The candidate paths of path-planning scenario}
\end{figure}

To simplify the path-planning process, we employed a widely used polynomial expression \cite{wang2023lane} to generate candidate paths. The representation of the candidate paths is as follows.
\begin{align} \label{eq12:gquadratic polynomial to generate candidate paths}
Y_{\text {ref }}=\epsilon_{0}+\epsilon_{1} X_{\text {ref }}+\epsilon_{2} X_{\text {ref }}^{2}+\epsilon_{3} X_{\text {ref }}^{3}+\epsilon_{4} X_{\text {ref }}^{4}+\epsilon_{5} X_{\text {ref }}^{5}
\end{align}%
where $Y_{\text {ref }}$ is the reference lateral position. $\epsilon_{0}, \epsilon_{1}, \epsilon_{2}, \epsilon_{3}$, $\epsilon_{4}$, and $\epsilon_{5}$ are the coefficients of the polynomial. In the path-planning process, the constraints, which include both the initial and terminal points of the lane-changing behavior, are designed as follows to determine these coefficients.

\begin{align}  \label{eq13:The constraint for the inital and terminal points of the lane-changing maneuver}
\left\{\begin{aligned}
 \left.Y_{\text {ref }}\right|_{X_{\text {ref },0}} =0,~    \left.\dot{Y}_{\text {ref }}\right|_{X_{\text {ref },0}}=0,~\left.\ddot{Y}_{\text {ref }}\right|_{X_{\text {ref },0}}=0 \\
 \left.Y_{\text {ref }}\right|_{X_{\text {ref },t}}=P,~\left.\dot{Y}_{\text {ref }}\right|_{X_{\text {ref },t}}=0,~\left.\ddot{Y}_{\text {ref }}\right|_{X_{\text {ref },t}}=0
\end{aligned}\right.
\end{align}
where $\left(X_{\text {ref },0}, Y_{\text {ref },0}\right)$ and $\left(X_{\text {ref },t}, Y_{\text {ref },t}\right)$ represent the initial and terminal position of the lane-changing maneuver along the reference path, respectively. 

Then by substituting Eq.(\ref{eq13:The constraint for the inital and terminal points of the lane-changing maneuver}) into Eq.(\ref{eq12:gquadratic polynomial to generate candidate paths}), the candidate paths can be further represented as follows.
\begin{align} \label{eq14:the expression of the reference paths combining the constraints}
Y_{\text {ref }}=\frac{10 F}{P^{3}} X_{\text {ref }}^{3}-\frac{15 F}{P^{4}} X_{\text {ref }}^{4}+\frac{6 F}{P^{5}} X_{\text {ref }}^{5}
\end{align} %
where $F$ and $P$ represent the lateral and longitudinal lane-changing distances, respectively. In this work, to facilitate system design, the longitudinal velocity of the AV is treated as constant during the lane-changing maneuver \cite{ji2016path}, which can be achieved through a Proportional-Integral algorithm in the motion-control module. Let $t_{c}$ denote the lane-changing duration, and hence the longitudinal distance of the lane-changing maneuver can be expressed as $P=V_{x} t_{c}$. Note that the mathematical expression for the bounds of $t_{c}$ is obtained by the constraint (\ref{eq15:The constraint for the safety distance and stability performance}):
\begin{align} \label{eq15:The constraint for the safety distance and stability performance}
\left\{\begin{aligned}
& d_{aovi }=V_{x, \text { rel }} t_{c}+\frac{1}{2} a_{x, \max } t_{c}^{2} \leq\left(\min \left\{d_{c}, d_{a}\right\}-s_{saf}\right)\\
&\beta_{\text {ref }} \leq\left|\tan ^{-1}(0.02 \mu g)\right| 
\end{aligned}\right.
\end{align}

Eq.(\ref{eq15:The constraint for the safety distance and stability performance}) is derived to ensure the collision avoidance and stability of the AV, where $d_{avoi }$ represents the collision-avoidance distance, and $d_{c}$ and $d_{a}$ are the available space for the lane-changing maneuver at the current and adjacent lanes, respectively. $s_{saf}$ indicates the minimum safety distance between vehicles, $a_{x, \max }$ is the maximum allowed deceleration rate, and $\mu$ is the road adhesion coefficient. $\beta_{\text {ref}}$ represents the vehicle sideslip angle at the reference path, which can serve as an indicator to characterize the AV's stability in vehicle dynamics \cite{liang2021distributed}. Specifically, the reference value of $\beta_{\text {ref}}$ can be calculated as follows.
\begin{align} \label{eq16: reference states along the candidate paths}
\quad \beta_{r e f}=\frac{\dot{Y}_{r e f}}{V_{x}}
\end{align} % 
where
\begin{align}
\dot{Y}_{\text {ref }}=\frac{30 F}{t_{c}^{3}} t^{2}-\frac{60 F}{t_{c}^{4}} t^{3}+\frac{30 F}{t_{c}^{5}} t^{4}
\end{align} 

A set of candidate paths as shown in Fig. \ref{Fig4:The candidate paths of path-planning scenario} are generated from Eq.(\ref{eq14:the expression of the reference paths combining the constraints}) with a time interval of 0.1\,s between $t_{c, \text { min }}$ and $t_{c, \text { max }}$, where 
Let $t_{c, \min }$ and $t_{c, \max }$ denote the lower and upper bounds for $t_c$, respectively, calculated from Eq.(\ref{eq15:The constraint for the safety distance and stability performance}).

\subsection{IRL-Based Optimal Trajectory Selection} 
\vspace{0.5em}

To make the behavior of AVs (characterized by their trajectories) understandable by HDVs, this work captures the preference of human drivers over multiple control objectives by introducing an IRL approach to learn skilled drivers' reward function for planning lane-changing paths. After finishing training, an optimal reference path would be selected to achieve a human-like tradeoff between these control objectives.

\vspace{0.5em} 
\noindent \emph{1) IRL Reward Construction}
\vspace{0.5em} \\
In this work, we generate human-like paths by learning the preferences of human drivers over various control objectives, including vehicle safety, driving comfort, and travel efficiency. Such preferences are characterized by a reward function represented as a linear combination of these objectives,  whereby the weights are learned via an IRL framework. The training data to the IRL framework involves the paths generated by skilled drivers on a driving simulator, because skilled drivers are more likely able to control the vehicles according to their preferences, while the actions of unskilled drivers may be subject to unexpected errors. 

Specifically, AV's safety is quantified by considering the stability of the vehicle dynamics and the potential collision risk. Driving comfort is indicated by the change rate of the yaw angle, while travel efficiency is represented by the lane-changing duration $t_{c}$. A vector of control objectives for a considered candidate path is constructed as follows: 
\begin{align} \label{eq18:feature vector of the performance indices in the IRL method}
H=\left[H_{sta}, H_{col}, H_{com}, H_{tra}\right]
\end{align} %
where $H_{sta},~H_{col},~H_{com}$ and $H_{tra}$ represent the control objectives for vehicle stability, collision risk, comfort, and travel efficiency, respectively, with the specific forms given as follows: 
\begin{align} 
\left\{\begin{array}{l}
\displaystyle H_{sta}=\frac{1}{\eta}\sum_{l=1}^{\eta} \beta_{l}^{2}, H_{com}=\frac{1}{\eta}\sum_{l=1}^{\eta} \dot{\varphi}_{\text {l}}^{2}, H_{tra}=\frac{t_{c}^{2}}{\eta}, \\
\displaystyle H_{col}=\frac{1}{\eta}\sum_{l=1}^{\eta} \sum_{m=1}^{n} \exp \Big[\sigma_{1}\left(Y_{\text{ref},l}-Y_{m,l}\right)^{2}+ \\ 
\quad\quad\quad\quad\sigma_{2}\left(X_{\text{ref},l}-X_{m,l}\right)^{2}\Big]
\end{array}\right.
\end{align} %
where we discretize the lane-changing duration $[t_{c,\min}, t_{c,\max}]$ into $\eta$ time instants with equal intervals of 0.1\,s, indexed by $l$, and $\dot{\varphi}_{\text {l}}$ represents the change rate of the yaw angle at the time instant $l$. 
 
The reward function of IRL is then constructed by a linear combination of the feature vector.
\begin{align} \label{eq21: reward of the IRL method}
r=\varpi^{T} H
\end{align} %
where $\varpi=\left[\varpi_{sta}, \varpi_{col}, \varpi_{com}, \varpi_{tra}\right]$.

\vspace{0.5em}
 \noindent \emph{2) IRL Training Process}
 \vspace{0.5em}

During the training process, we utilize the driving behavior of skilled drivers as expert experience to optimize the feature weight coefficients $\varpi$ in Eq.(\ref{eq21: reward of the IRL method}).  Note that we adopt a maximum entropy method \cite{ziebart2008maximum} for training the IRL model. The objective of such method is to maximize the likelihood of the driver's trajectory $\lambda_{q} \in \Theta,(q=1,2, \ldots, Q)$. %

\begin{align} \label{eq22: optization objective of the  maximum entropy method}
\max _{\varpi} \sum_{\lambda \in \Theta} \log P(\lambda \mid \varpi)
\end{align} 

\begin{align} 
P(\lambda \mid \varpi)=\frac{e^{\varpi^{T} H_{\lambda}}}{\sum_{j=1}^{S} e^{\sigma^{T} H_{\bar{\lambda}_{j}}}}
\end{align} 
where $Q$ indicates the total driver's trajectories collected in the tests. $\tilde{\lambda}_{j}$ is the candidate path generated by the polynomial expression Eq.(\ref{eq14:the expression of the reference paths combining the constraints}). $S$ and $H_{\lambda}$ represent the number of candidate paths and feature vector of the driver trajectory, respectively. To guarantee the effectiveness of the training data, the initial state when generating the trajectory $\tilde{\lambda}_{j}$ is the same as that of driver trajectory $\lambda$. 

The objective function of the IRL framework can be written as follows.
\begin{align} \label{eq24: The mathematical expression for the objective function}
\Omega(\varpi)=\sum_{\lambda \in \Theta}\left(\varpi^{T} H_{\lambda}-\log \sum_{j=1}^{S} e^{\varpi^{T} H_{\tilde{\lambda}_{j}}}\right)
\end{align}
The gradient of $\Omega(\varpi)$ is computed as:
\begin{align}\label{eq25: gradient}
\nabla_{\varpi} \Omega(\varpi) & =\sum_{\lambda \in \Theta}\left(H_{\lambda}-\sum_{j=1}^{S} \frac{e^{\varpi^{T} H_{\bar{\lambda}_{j}}}}{\sum_{j=1}^{S} e^{\varpi^{T} H_{\tilde{\lambda}_{j}}}} H_{\tilde{\lambda}_{j}}\right) \\ \nonumber
& =\sum_{\lambda \in \Theta}\left(H_{\lambda}-\sum_{j=1}^{S} P(\tilde{\lambda} \mid \varpi) H_{\tilde{\lambda}_{j}}\right)
\end{align}

Eq.(\ref{eq24: The mathematical expression for the objective function}) could reflect the difference between the driver trajectory and candidate paths. It is employed to iterate and update the feature weight vector $\varpi$ using the gradient ascent method Eq.(\ref{eq25: gradient}). The IRL algorithm flow is presented in Algorithm 2. After completing the training, the feature weight vector is determined. Hence, the AV's path-planning ability can emulate a skilled driver to effectively balance different control objectives in high-speed cruising scenarios. In the IRL framework, the learning rate $\rho$ and episodes $E$ can impact the training effectiveness. By comparing the human-like driving results with different settings, $\rho=0.08$ and $E=150$.

\begin{algorithm}[ht]
	\caption{IRL algorithm}
	\renewcommand{\algorithmicrequire}{\textbf{Input:}}
	\renewcommand{\algorithmicensure}{\textbf{Initialize:}}
	\begin{algorithmic}[1]
		\REQUIRE  Trajectories of skilled driver obtained in the driving simulator $\lambda_{q} \in \Theta,(q=1,2, \ldots, Q)$, learning rate $\rho$, and episodes $E$.
		\ENSURE Feature weight vector $\varpi\leftarrow \varpi_0$.
        \STATE Calculate the feature vector of driver trajectories $\sum_{q=1}^{Q} H_{\lambda_{q}}$.
		\FOR{each $\lambda_{q} \in \Theta,(q=1,2, \ldots, Q)$,}
		\STATE  Generate candidate path $\tilde{\lambda}_{j}$ according to potential selections of lane-changing duration $t_{c}$, with the same initial state as 
 $\lambda_{q}$.
		\FOR{each $\tilde{\lambda}_{j}$,}
        \STATE Based on the vehicle states on the candidate paths, calculate the feature vector $H_{\bar{\lambda}_{j}}$.
        \STATE Store $\tilde{\lambda}_{j}$ and $H_{\tilde{\lambda}_{j}}$ to the buffer $B^{\prime}$.
		\ENDFOR
		\ENDFOR
      \FOR{each episode,}
       \STATE Use the variables in buffer to calculate the gradient $\nabla_{\varpi} \Omega(\varpi)$. 
     \STATE Update the IRL optimization parameter $\varpi=\varpi+\rho \nabla_{\varpi}$.
      \ENDFOR
  \STATE Then $\varpi^{\prime} \leftarrow \varpi$, where $\varpi^{\prime}$ is the optimized feature weight vector.
	\end{algorithmic}
\end{algorithm}
\section{Motion-control Module Design via MPC Method}
In this section, we develop an MPC-based tracking controller to implement the motion control of potential lane-keeping and lane-changing behaviors, in order to accurately track the planned path while respecting constraints on physical dynamics and safety.
\subsection{Lane-Keeping Motion Control}  
\vspace{0.5em}
The lane-keeping motion-control module aims to track the reference acceleration/deceleration signal $a_{x, \text { ref }}$ calculated in the behavioral decision module, while ensuring high-speed cruising of the ego AV. This is achieved via a MPC-based controller with an embedded quadratic programming (QP) problem described as follows. Note that the system state-space equation adopts the vehicle longitudinal dynamics model and can be derived from Eq.(\ref{eq4:state-space equation}) \cite{vicente2020linear}.

\begin{subequations}
    \begin{align} 
\min \quad & J_{keeping}= \sum_{t=1}^{N_{p}}\Big(\left\|V_{x}(k+t \mid k)-V_{x, \max }\right\|_{P_{1}}^{2}\notag \\ &+ 
\left\|a_{x}(k+t \mid k)-a_{x,  r e f}\right\|_{R_{1}}^{2}\Big) \label{eq26: The objective function for the lane-keep maneuver.}\\
\rm{s.t.}\quad & a_{x, \min } \leq a_{x}(k) \leq a_{x, \max } \label{eq:motion_keep_con1}\\
& V_{x, \min } \leq V_{x}(k) \leq V_{x, \max } \label{eq:motion_keep_con2} \\
& X_{p r e}(k)-X_{e v}(k) \geq s_{saf} \label{eq:motion_keep_con3}
\end{align}
\end{subequations}
where the objective function Eq.(\ref{eq26: The objective function for the lane-keep maneuver.}) minimizes the total costs over a prediction horizon $N_p$, including (i) the deviation from the maximum speed (the first term) and (ii) the tracking error between the actual acceleration rate and the reference acceleration rate $a_{x,ref}$ given by the decision-making module (the second term), weighted by constants $P_{1}$ and $R_{1}$. 
The specific values of $P_{1}$ and $R_{1}$ are determined by the DRL agent in the behavioral decision-making module. Constraints (\ref{eq:motion_keep_con1}) and (\ref{eq:motion_keep_con2}) impose physical bounds on the acceleration and velocity at each decision step. Constraint (\ref{eq:motion_keep_con3}) ensures that the spacing between the ego AV and the preceding HDV is lower bounded by the safe spacing. Notice that here we assume that the following HDV will not actively collide with the AV.  

\subsection{Lane-Changing Motion Control}  
\vspace{0.5em}
The lane-changing motion-control module devises an MPC-based controller to track the reference path generated by the path-planning module, while ensuring the AV's lateral stability \cite{zhang2019non}. The embedded optimization problem of the MPC can be represented as follows. The system state-space equation employs the vehicle lateral dynamics model \cite{ji2016path}.
\begin{subequations}
    \begin{align}
\min\quad & J_{changing}=\sum_{t=1}^{N_{p}}\Big(\left\|\bar{\zeta}(k+t \mid k)-\bar{\zeta}_{\text {pre }}\right\|_{P_{2}}^{2} \notag \\ 
&\quad\quad\quad\quad\quad +\left\|\delta_{f}(k+t \mid k)\right\|_{R_{2}}^{2}\Big) \label{eq27: The objective function for the lane-k maneuver.} \\
\rm{s.t.}\quad & \delta_{f, \min } \leq \delta_{f}(k) \leq \delta_{f, \max } \label{eq:motion_change_con1}\\
&|\gamma| \leq \frac{\mu g}{V_{x}} \label{eq:motion_change_con2} \\
& \left|\alpha_{f}\right| \leq \arctan \frac{\mu M g L_{r}}{2 K_{f}\left(L_{f}+L_{r}\right)} \label{eq:motion_change_con3}
\\
&\left|\alpha_{r}\right| \leq \arctan \frac{\mu M g L_{f}}{2 K_{r}\left(L_{f}+L_{r}\right)} \label{eq:motion_change_con4}
\end{align}%
\end{subequations}
where $\bar{\zeta}=[\varphi, Y]^{T}$, and $\bar{\zeta}_{ref}=\left[\bar{\varphi}_{\text {ref }}, Y_{\text {ref }}\right]^{T}$. The objective function Eq.(\ref{eq27: The objective function for the lane-k maneuver.}) includes (i) the deviation from the reference trajectories (the first term) and (ii) the cost of the control inputs (the second term), weighted by positive-definite matrices $P_{2}\in\mathbb{R}^{2\times2}$ and $R_{2}\in\mathbb{R}$. 
The specific values of $P_{2}$ and $R_{2}$ are determined by the behavioral decision-making module, where
$P_{2}=\operatorname{diag}\left(P_{21}, P_{22}\right)$. $\bar{\varphi}_{\text {ref }}=0$ is set to guarantee the AV's heading angle is zero after completing the lane-changing maneuver. Constraint (\ref{eq:motion_change_con1}) imposes physical bounds on the front-wheel steering input. Constraints (\ref{eq:motion_change_con2}), (\ref{eq:motion_change_con3}), and (\ref{eq:motion_change_con4}) ensure the AV's lateral handling stability \cite{zhang2019non}.

\section{Simulation} \label{Simulation Tests}
In this section, simulations are conducted to verify the effectiveness of the proposed method using a Simulink/Python joint platform. The vehicle model in Section II and the motion-control module are constructed in Simulink, while the behavioral decision-making and path-planning modules are implemented with Python.

\begin{table}[ht] 
	\begin{center}
		\caption{Parameters Setting}
		\begin{tabular}{ c  c }
			\hline
   Parameters & Bootstrapped DQN \\
			\toprule  
		 Learning rate & $1 \mathrm{e}-4$ \\
Reward discount & 0.95 \\
Batch size & 64 \\
Memory capacity & 15000 \\
Target replace & 500 \\
Network layers & 3 \\
Number of neurons & {$[128,64,64]$} \\
			\bottomrule 
			\hline 
		\end{tabular}
  \label{tab4: The parameters for the decision-making RL network}
	\end{center}
\end{table}

\subsection{Training Process of the Integrated Framework}
The integrated framework requires the training of the feature weight vector $\varpi$ and the behavioral decision-making policy, which is performed in a sequential manner. 
First, using collected data of skilled human drivers, the IRL method is trained based on Algorithm 2 to obtain the feature weight vector $\varpi$. Then, Algorithm 1 is employed to train an AV navigating in a high-speed cruising scenario by integrating the behavioral decision-making, path-planning, and motion-control modules. After tuning in the training progress, the hyperparameters of the bootstrapped DQN are given in Table \ref{tab4: The parameters for the decision-making RL network}. Note that 
during the training process, the initial positions and velocities of HDVs are randomly set. Meanwhile, the Long Short-Term Memory network is employed to obtain the driving behavior of the leading HDVs from the historical data.

Furthermore, to simplify the system design, we set a benchmark for the training of control weights in the motion-control module. For lane-keeping control (Eq.(\ref{eq26: The objective function for the lane-keep maneuver.})), to prioritize the optimization objective of tracking the reference acceleration signal, $R_{1}=4 P_{1}$ is set as a benchmark. As for the lane-changing control (Eq.(\ref{eq27: The objective function for the lane-k maneuver.})), considering the priority of path-tracking performance, we set $P_{22}=10 P_{21}$. After completing the training, the RL agent in the behavioral decision-making module would output the specific values of control weights.

\begin{figure}[ht]
	\centering
	\includegraphics[width=3.4in]{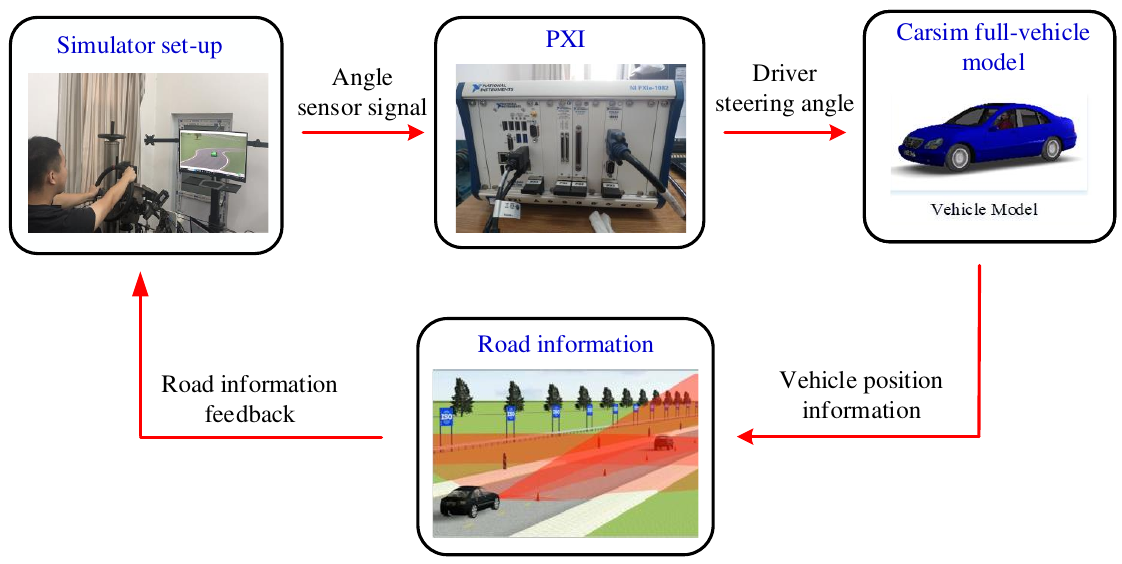}
	\caption{Driving simulator platform.}
 \label{ fig 5:Driving simulator platform.}
\end{figure}%

\subsection{Verification of Human-Like Path-Planning}

In the path-planning module, we introduce the IRL method to learn skilled drivers' reward function for planning lane-changing paths. The driving simulator, as depicted in Fig. \ref{ fig 5:Driving simulator platform.}, is employed to collect driver behavior data. More details regarding the driving simulator can be found in \cite{liang2023robust}. 
We invite skilled drivers to participate in the tests, who will get familiar with the testing equipment in advance to ensure the accuracy of the test data.

Through these experiments, we construct a dataset comprising 50 drivers' lane-changing trajectories under different test conditions. This dataset is utilized to update the IRL weight vector $\varpi$. After finishing the training, we employ a set of 10 driver trajectories to evaluate the learning performance. Here, the feature vector $H$ in Eq.(\ref{eq18:feature vector of the performance indices in the IRL method}) is calculated to provide a comparison of human-like path-planning performance. The specific indices from IRL's learning results and driver's trajectories are presented in Table \ref{tab4: test results with different control strategies}. The average learning errors for different performance indices, including vehicle safety, driving comfort, and travel efficiency, in ten tests are $6.92 \%, 3.36 \%, 6.14 \%$, and $0.83 \%$, respectively. These results demonstrate the effectiveness of the proposed IRL method in acquiring preferences of human drivers, thereby balancing different control objectives during lane-changing maneuvers.

\begin{table}[ht!]
\centering
\caption{Test Results of Human-like Performance}
\begin{tabular}{cc|cccc}
\Xhline{1.5pt}
\multicolumn{2}{c|}{\begin{tabular}[c]{@{}c@{}}Performance\\ index\end{tabular}} & \begin{tabular}[c]{@{}c@{}}Stability\\ performance\end{tabular} & \begin{tabular}[c]{@{}c@{}}Potential\\ collision\end{tabular} & \begin{tabular}[c]{@{}c@{}}Driving\\ comfort\end{tabular} & \begin{tabular}[c]{@{}c@{}}Travel\\ efficiency\end{tabular} \\ \hline
\multirow{2}{*}{\begin{tabular}[c]{@{}c@{}}Test\\ case1\end{tabular}}   & IRL      & 0.000395   & 167.571     & 22.524                   & 0.105                                  \\ \cline{2-2}     & Driver  & 0.000416       & 163.312            & 30.104           & 0.095                    \\ \hline
\multirow{2}{*}{\begin{tabular}[c]{@{}c@{}}Test\\ case2\end{tabular}}   &  IRL     & 0.000204                                                         & 133.042                                                          & 16.958                                                       & 0.12                                                        \\ \cline{2-2}     & Driver  & 0.000175                                                          &   141.730                                                        & 19.654                                                       &  0.13                                                       \\ \hline
\multirow{2}{*}{\begin{tabular}[c]{@{}c@{}}Test\\ case3\end{tabular}}   & IRL      & 0.000257                                                          & 121.956                                                        & 20.652                                                       & 0.115                                                        \\ \cline{2-2}     & Driver  & 0.000325                                                          & 134.513                                                          & 18.293                                                       & 0.115                                                        \\ \hline
\multirow{2}{*}{\begin{tabular}[c]{@{}c@{}}Test\\ case4\end{tabular}}   & IRL     & 0.000181                                                          & 152.038                                                          & 19.024                                                       & 0.13                                                        \\ \cline{2-2} & Driver  & 0.000258                                                          & 160.642                                                          & 21.924                                                       & 0.125                                                        \\ \hline
\multirow{2}{*}{\begin{tabular}[c]{@{}c@{}}Test\\ case5\end{tabular}}   & IRL       & 0.000257                                                          & 118.793                                                          & 17.462 
                                                 & 0.13
                                                 \\ 
 \cline{2-2}     & Driver  & 0.000218                                                         & 115.923                                                          & 18.942                                                       & 0.13                                                        \\ \hline
\multirow{2}{*}{\begin{tabular}[c]{@{}c@{}}Test\\ case6\end{tabular}}   &  IRL      & 0.000451                                                         & 169.723                                                         & 25.863                                                       & 0.11                                                        \\ \cline{2-2}     & Driver  & 0.000497                                                          & 181.543                                                          & 28.727                                                       & 0.11                                                       \\ \hline
\multirow{2}{*}{\begin{tabular}[c]{@{}c@{}}Test\\ case7\end{tabular}}   & IRL      & 0.000263                                                          & 118.192                                                          & 21.038                                                       & 0.13                                                        \\ \cline{2-2}     &  Driver  & 0.000291                                                         & 123.321                                                          & 19.526                                                       & 0.14                                                        \\ \hline
\multirow{2}{*}{\begin{tabular}[c]{@{}c@{}}Test\\ case8\end{tabular}}   & IRL      & 0.000236                                                          & 143.160                                                          & 22.481                                                       & 0.125                                                        \\ \cline{2-2}&  Driver  & 0.000248                                                         & 138.379                                                          & 21.017                                                       & 0.115                                                        \\ \hline
\multirow{2}{*}{\begin{tabular}[c]{@{}c@{}}Test\\ case9\end{tabular}}   & IRL      & 0.000256                                                          & 121.815                                                          & 17.556                                                       & 0.135                                                        \\ \cline{2-2}& Driver  & 0.000317                                                          &   113.679                                                       & 20.571                                                    & 0.14                                                       \\ \hline
\multirow{2}{*}{\begin{tabular}[c]{@{}c@{}}Test\\ case10\end{tabular}}  & IRL      & 0.000421                                                          & 164.552                                                          & 25.471                                                       & 0.10                                                        \\ \cline{2-2} &  Driver & 0.000393                                                         & 179.727                                                          & 23.954                                                       & 0.11                                                        \\ \Xhline{1.5pt}
\end{tabular}
\label{tab4: test results with different control strategies}
\end{table}

\subsection{Verification of the Integrated Framework}
\vspace{0.5em}
\noindent \emph{1) Overall Performance Analysis}
\vspace{0.5em}\\
Simulations are conducted to demonstrate the value of the integrated framework in enhancing the overall performance of autonomous driving. Specifically, we compare the resulting performance of the proposed integrated framework with that of a sequential framework that involves separate training and execution of the three fundamental modules, i.e., behavioral decision-making, path-planning, and motion-control. Note that to decouple the training of fundamental modules in the sequential framework, we train the behavioral decision-making module by assuming a simplified path planning procedure with a constant lane-changing execution time (i.e., 2 seconds) and motion control with pre-defined constant control weights.   

\begin{figure}[ht]
	\centering
	\includegraphics[width=3.7in]{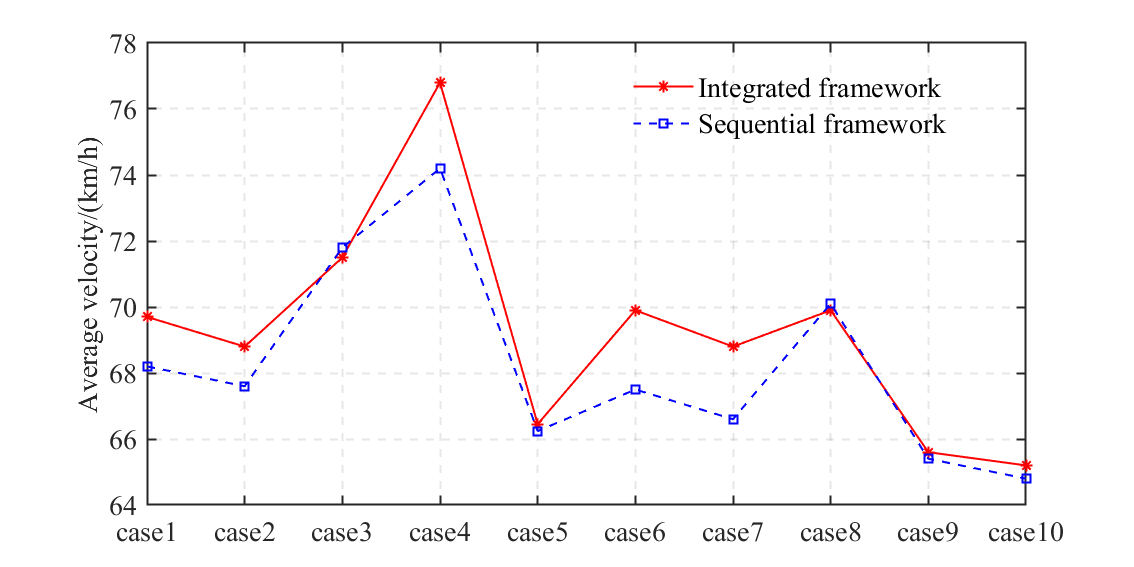}
	\caption{Average velocity  with different frameworks.}
 \label{ fig 6: Average velocity  with different frameworks.}
\end{figure}

\begin{figure}[ht]
	\centering
	\includegraphics[width=3.7in]{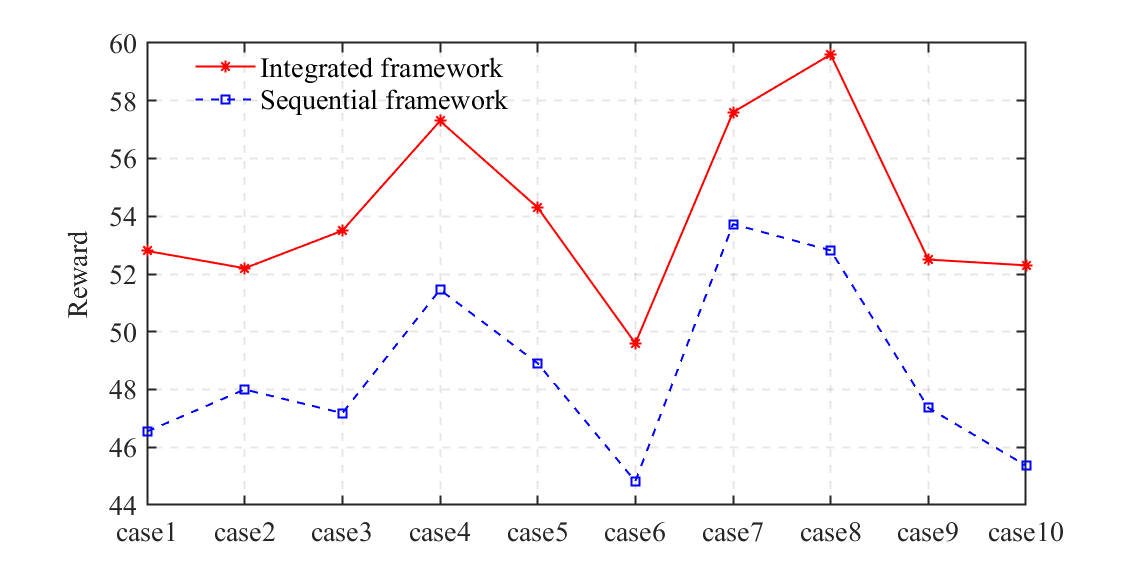}
	\caption{Reward with different frameworks.}
 \label{ fig 7: Reward with different frameworks.}
\end{figure}

After completing the training for these two frameworks, ten test cases with different initial positions and velocities of HDVs are conducted to verify the control performance. As shown in Figs. \ref{ fig 6: Average velocity  with different frameworks.} and \ref{ fig 7: Reward with different frameworks.}, it is clear that the proposed integrated framework outperforms the sequential framework in both the average AV velocity and reward. Overall, the total average velocity and reward with the integrated framework for ten test cases can be improved by $2.12\%$ and $10.25\%$, respectively. Although the average velocity of the AV with the sequential framework is slightly higher than that of the integrated framework in test cases 3 and 8, the integrated framework performs significantly better in ensuring a high reward. This is because the sequential framework cannot make optimal decisions from a global perspective due to the lack of coordination between these fundamental modules. The results demonstrate that the integrated framework can enhance high-speed cruising performance while considering collision avoidance, the interpretability of AVs’ trajectories, and the feasibility of AV motion.

\begin{table}[ht]
	\begin{center}
		\caption{Evaluating Indicators for Longitudinal Dynamics Performance}
		\begin{tabular}{c c}
			\hline
			\toprule  
			Description & Math expression \\
			\midrule  
	AV's longitudinal tracking \! \!\!\!\!& \!\!\!$E_{1}\!=\!\frac{1}{\mathrm{~K}\left|a_{x, \max }\right|} \!\!\sum_{t=1}^{\mathrm{K}}\left|a_{x, t}\!\!-\!\!a_{x, \text { ref }}\right|$ \\
High-speed cruising\!\!\! &\!\!\! $E_{2}\!=\!\frac{1}{\mathrm{~K}\left|V_{x, \max }\right|} \!\!\sum_{t=1}^{\mathrm{K}}\left|V_{x, t}\!\!-\!\!V_{x, \max }\right|$ \\
 Control effort\!\!\! &\!\!\! $E_{3}\!=\!\frac{1}{\mathrm{~K}\left|a_{x, \max }\right|} \!\!\sum_{t=1}^{\mathrm{K}}\left|a_{x, t}\right|$ \\
			\bottomrule 
			\hline 
		\end{tabular}
  \label{ tab5:Evaluating indicators with different settings for lane-keeping control}
	\end{center}
\end{table}

\begin{table}[ht]
	\begin{center}
		\caption{Evaluating Indicators for Lateral Dynamics Performance}
		\begin{tabular}{c c}
			\hline
			\toprule  
			Description & Math expression \\
			\midrule  
	 Path-tracking accuracy &$\bar{E}_{1}=\frac{1}{\mathrm{~K}\left|Y_{\max }\right|} \sum_{t=1}^{\mathrm{K}}\left|Y_{t}-Y_{\text {ref }}\right|$\\
 Stability state & $\bar{E}_{2}=\frac{1}{\mathrm{~K} \mid \beta_{\max }} \sum_{t=1}^{\mathrm{K}}\left|\beta_{t}\right|$\\
control effort & $\quad \bar{E}_{3}=\frac{1}{\mathrm{~K}\left|\delta_{f, \text { max }}\right|} \sum_{t=1}^{\mathrm{K}}\left|\delta_{f, t}\right|$\\
			\bottomrule 
			\hline 
		\end{tabular}
  \label{ tab6:Evaluating indicators with different settings for lane-changing control}
	\end{center}
\end{table}

\vspace{0.5em}
\noindent \emph{2) Value of integrating the motion-control module}
\vspace{0.5em}

\begin{figure}[ht]
	\centering
 \begin{subfigure}{0.5\textwidth}     
	\includegraphics[width=3.55in]{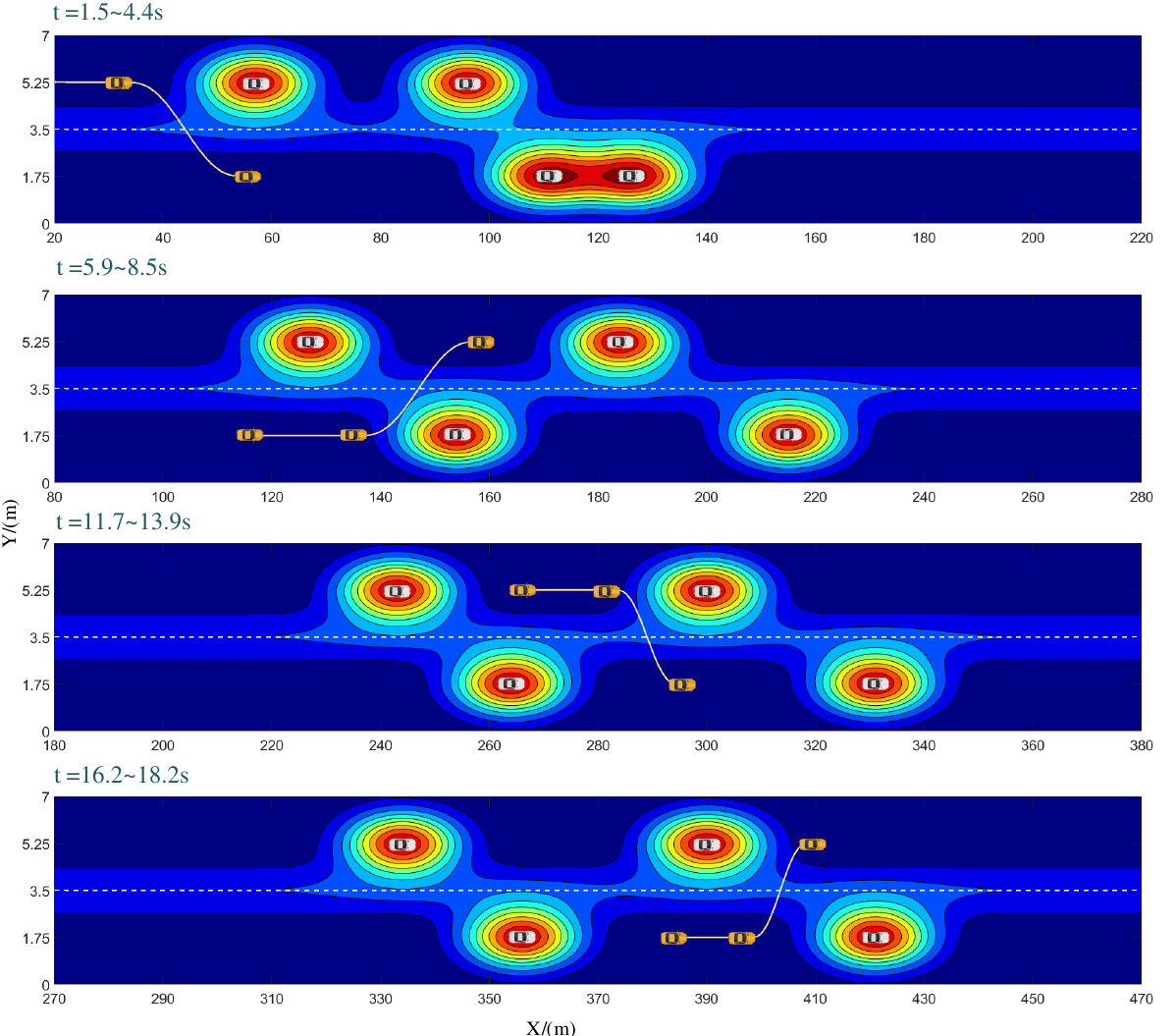}
	\caption{The trajectory of the ego AV.} \label{ fig 8:The trajectory of the ego AV in test 1.}
  \end{subfigure}
 \begin{subfigure}{0.5\textwidth}
     \includegraphics[width=3.7in]{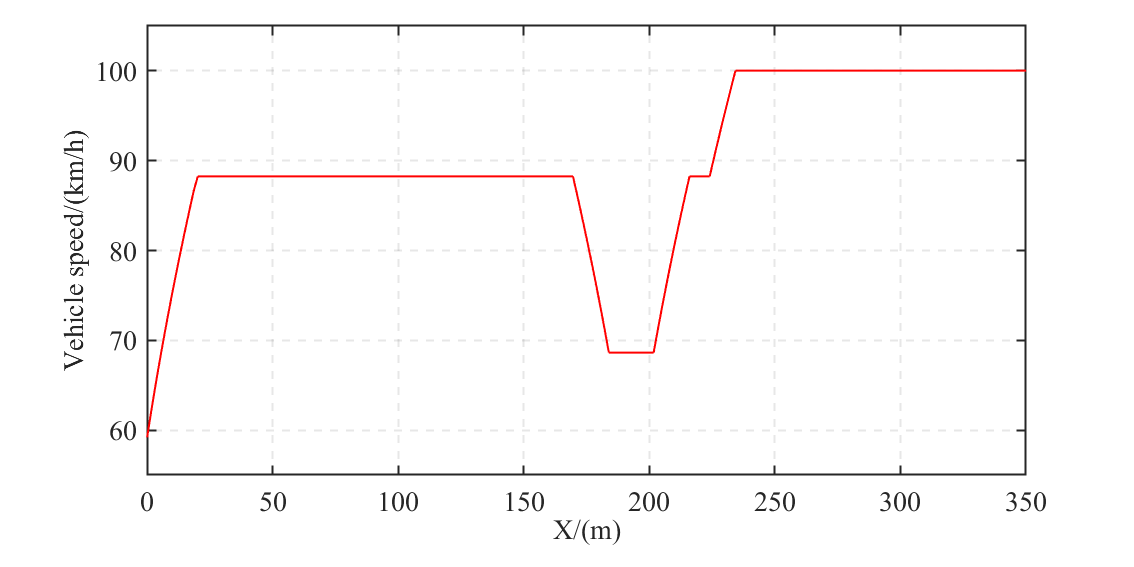}
	\caption{The speed of the ego AV.}
 \label{fig9:The speed of the ego AV.}
 \end{subfigure}
 \caption{ The continuous overtaking maneuver scenario under the integrated framework. The ego AV performs lane changes at $t=1.5$\,s, $t=5.9$\,s, $t=11.7$\,s and $t=16.2$\,s to achieve a higher speed (i.e., $88\,\rm{km/hr},~88\,\rm{km/h},~100\,\rm{km/h}$, and $100\,\rm{km/h}$, respectively) while keeping safe distance with the front HDVs. The resulting lane-changing duration ranges from $2 \mathrm{~s}$ to $3 \mathrm{~s}$ during high-speed cruising, consistent with the typical driving profile of a skilled driver. }
 \label{fig:continuous overtaking maneuver scenario}
\end{figure}

We specifically highlight the value of integrating the motion-control module by dynamically setting up control weights, which has not been considered in the existing literature. To this end, we compare our integrated framework with a semi-integrated framework that only incorporates the path-planning module, while the control weights in the motion-control module are assumed to be pre-defined constants.  
For both the integrated and semi-integrated frameworks, we simulate a continuous overtaking maneuver scenario, which provides a comprehensive test case to better assess the longitudinal and lateral driving performance of AVs. Specifically, the continuous overtaking maneuver scenario under the integrated framework is shown in Fig. \ref{fig:continuous overtaking maneuver scenario}. 

\begin{figure}[ht]
	\centering
	\includegraphics[width=3.6 in]{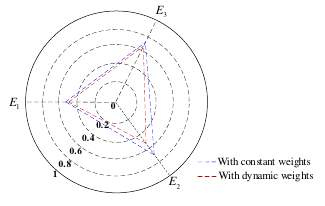}
	\caption{Evaluating indicators for longitudinal dynamics performance in test 1.} \label{fig10:evaluating indicator 1}
\end{figure}

\begin{figure}[ht] 
	\centering
	\includegraphics[width=3.6 in]{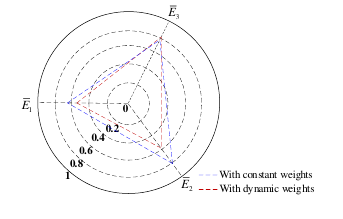}
	\caption{Evaluating indicators for lateral dynamics performance in test 1.}\label{ fig 11:Evaluating indicators 2}
\end{figure}%

\begin{figure}[ht]
	\centering
	\includegraphics[width=3.8 in]{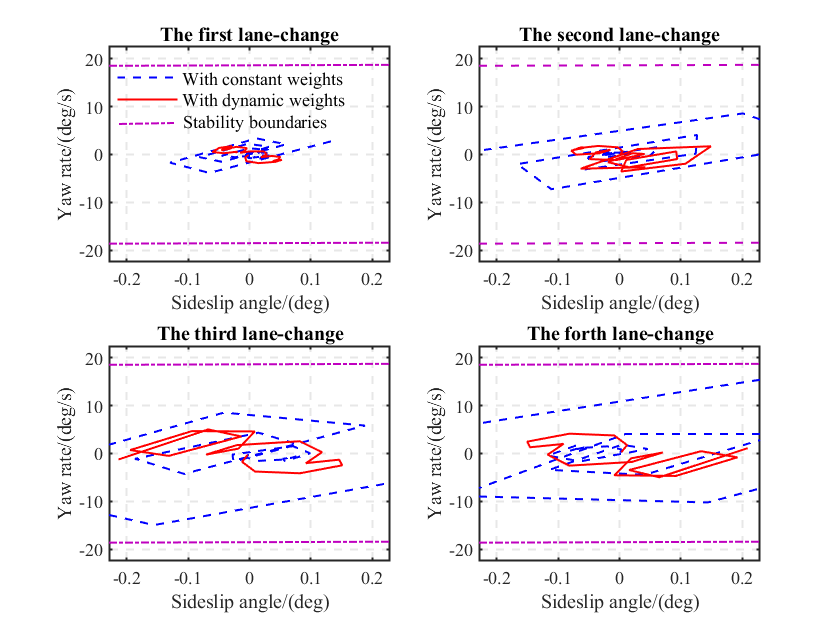}
	\caption{The lateral motion phase plane of the ego AV with and without the constraint in test 1.}
  \label{fig12:The lateral motion phase plane of the ego AV}
\end{figure}

The value of integrating the motion-control module is illustrated by evaluating the performance indicators summarized in Tables \ref{ tab5:Evaluating indicators with different settings for lane-keeping control} and \ref{ tab6:Evaluating indicators with different settings for lane-changing control}, respectively, for the longitudinal and lateral control of AVs. The comparison between the integrated framework (i.e., with dynamic weights) and the semi-integrated framework (i.e., with constant weights) is shown in Figs. \ref{fig10:evaluating indicator 1} and \ref{ fig 11:Evaluating indicators 2}, whereby it is clear that the ego AV exhibits better high-speed cruising performance and stability. This can be attributed to the DRL agent's ability to dynamically adjust control weights, thereby ensuring optimal control performance in dynamic driving conditions. From Fig. \ref{ fig 11:Evaluating indicators 2}, although the semi-integrated framework with constant control weights consumes slightly less control effort, it performs worse in ensuring path-tracking and stability.
 
In addition, we assess the stability performance of the ego AV using the $\beta{\rm{-}}\gamma$ phase plane \cite{liang2022mas}. As depicted in Fig. \ref{fig12:The lateral motion phase plane of the ego AV}, when introducing the integrated framework with dynamic control weights, the ego AV runs within a smaller region, indicating a more stable state. As the third and fourth lane-changing maneuvers are executed at higher speeds, the stability state of the ego AV deteriorates when using constant control weights in the motion-control module. In contrast, the proposed integrated framework reliably ensures the AV's stability.

\subsection{Value of the Bootstrapped DQN for Training the Behavioral Decision-Making Module}
\vspace{0.5em}

\begin{figure}[ht]
	\centering
	\includegraphics[width=3.7in]{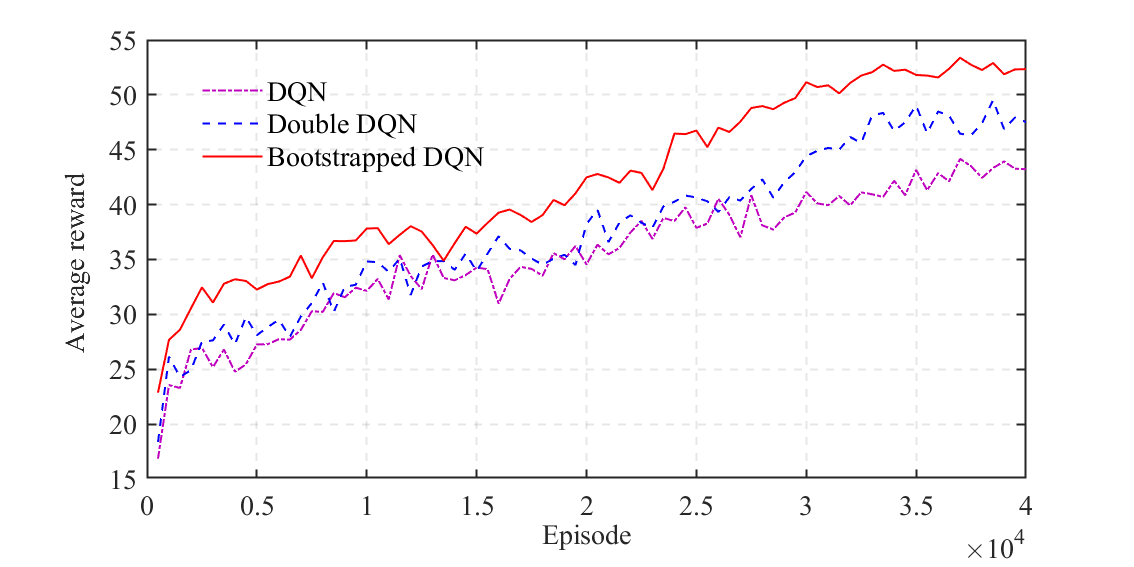}
	\caption{Average reward with different strategies.}
 \label{ fig 15:Average reward with different strategies.}
\end{figure}

\begin{figure}[ht]
	\centering
	\includegraphics[width=3.7in]{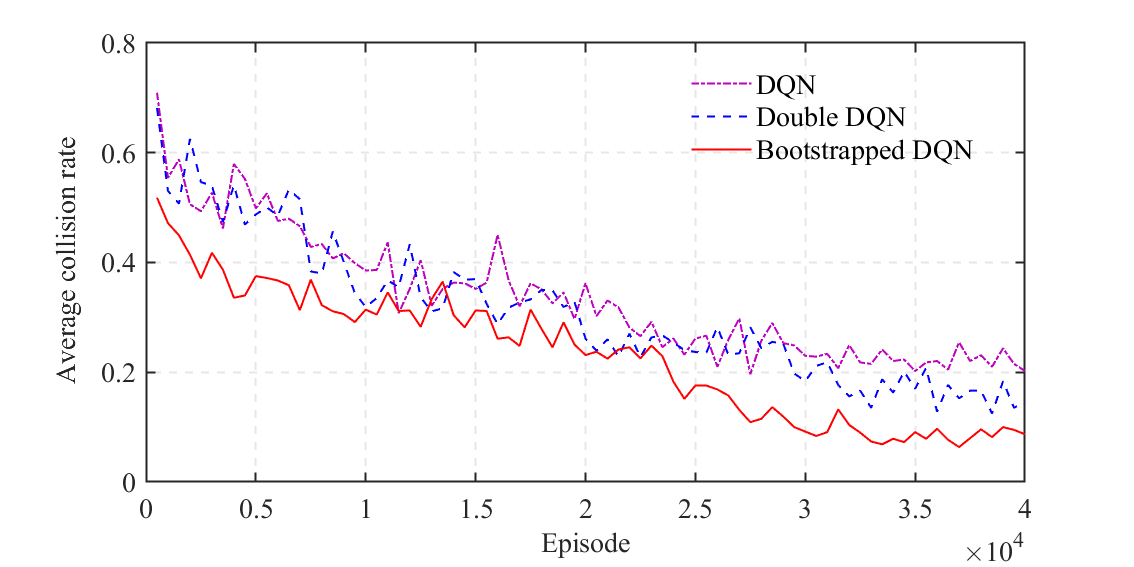}
	\caption{Average collision rate with different strategies.}
 \label{ fig 16:Average collision rate with different strategies.}
\end{figure}

\noindent \emph{1) Comparison of Different DRL Strategies}
\vspace{0.5em}\\
We demonstrate the value of using the Bootstrapped DQN for training the behavioral decision-making module by comparing it with other widely used DRL algorithms, including DQN and double DQN, in the same simulation environment as the proposed framework. As illustrated in Figs. \ref{ fig 15:Average reward with different strategies.} and \ref{ fig 16:Average collision rate with different strategies.}, the bootstrapped DQN outperforms other strategies in both the average reward and the collision rate. Specifically, the bootstrapped DQN exhibits the potential to achieve a maximum reduction of $62.31 \%$ and $43.59 \%$ in the average collision rate compared to the DQN and double DQN, respectively. These results indicate the ability of the bootstrapped DQN to enhance the training of the behavioral decision-making module in the complex integrated framework proposed in this work.

\vspace{0.5em}
\noindent \emph{2) Sensitivity Analysis on the Number of Head Networks in the Bootstrapped DQN}
\vspace{0.5em}

\begin{figure}[ht]
	\centering
	\includegraphics[width=3.7in]{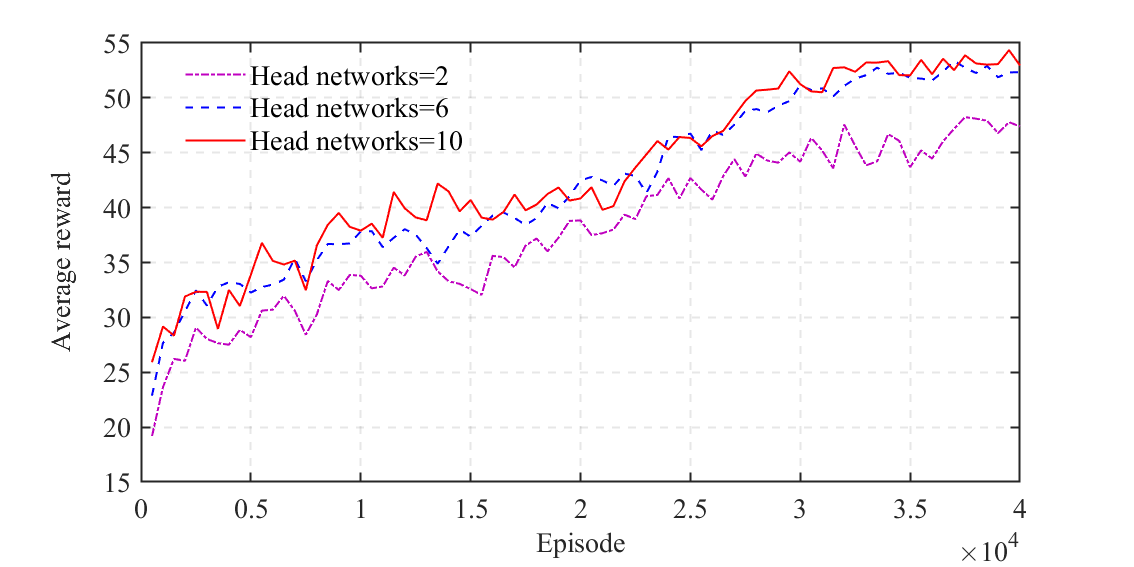}
	\caption{Average reward with different head networks.}
 \label{ fig 13:Average reward with different head networks.}
\end{figure}

\begin{figure}[ht]
	\centering
	\includegraphics[width=3.7in]{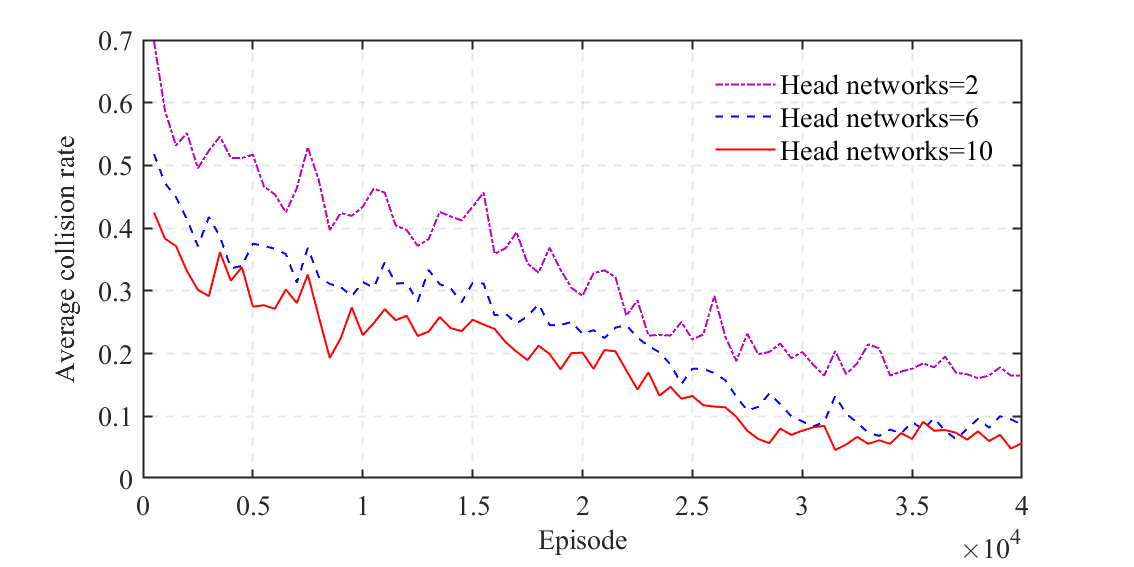}
	\caption{Average collision rate with different head networks.}
 \label{ fig 14:Average collision rate with different head networks.}
\end{figure}

We perform a sensitivity analysis on the number of head networks in the bootstrapped DQN. As shown in Figs. \ref{ fig 13:Average reward with different head networks.} and \ref{ fig 14:Average collision rate with different head networks.}, the bootstrapped DQN results in a higher average reward and a lower collision rate when employing more head networks. This can be attributed to the enhancement of the exploration capability with the increasing number of head networks. However, the marginal benefits of having more head networks diminish when the number of head networks reaches 6. This is because an adequate number of head networks could guarantee the precision of the uncertain Q-value estimation. Since more head networks consume more computing resources, we select the number of head networks as 6 in this work. 

\section{CONCLUSION}
In this study, we present an autonomous driving framework to systematically enhance the safe driving capabilities of AVs in high-speed cruising scenarios by integrating behavioral decision-making, path-planning, and motion-control modules. We further improve the interpretability of AV driving decisions by learning the reward function of skilled drivers for planning lane-changing paths in the path-planning module. Taking into account the complexity of the integrated framework, we introduce a bootstrapped DQN to enhance the deep-exploration ability. The DRL agent would adaptively choose between potential lane-keeping and lane-changing maneuvers, while generating control weights for the MPC-based motion-control module. Our simulation results indicate that the proposed integrated framework can effectively guide AVs to ensure high-speed cruising performance while avoiding collisions. In comparison to the conventional sequential framework, the average reward in test cases can be improved by $10.25\%$ by the proposed method. 

This paper opens several interesting directions for future work. First, we would like to extend our framework to enable cooperative control in connected and autonomous vehicle systems. Second, it would be interesting to improve the generalizability of the proposed framework to various driving scenarios (e.g., mandatory lane changes at highway bottlenecks, adverse weather conditions, etc.).

\bibliographystyle{IEEEtran}

\begin{thebibliography}{10}
\providecommand{\url}[1]{#1}
\csname url@samestyle\endcsname
\providecommand{\newblock}{\relax}
\providecommand{\bibinfo}[2]{#2}
\providecommand{\BIBentrySTDinterwordspacing}{\spaceskip=0pt\relax}
\providecommand{\BIBentryALTinterwordstretchfactor}{4}
\providecommand{\BIBentryALTinterwordspacing}{\spaceskip=\fontdimen2\font plus
\BIBentryALTinterwordstretchfactor\fontdimen3\font minus \fontdimen4\font\relax}
\providecommand{\BIBforeignlanguage}[2]{{%
\expandafter\ifx\csname l@#1\endcsname\relax
\typeout{** WARNING: IEEEtran.bst: No hyphenation pattern has been}%
\typeout{** loaded for the language `#1'. Using the pattern for}%
\typeout{** the default language instead.}%
\else
\language=\csname l@#1\endcsname
\fi
#2}}
\providecommand{\BIBdecl}{\relax}
\BIBdecl

\bibitem{zhao2020enhanced}
L.~Zhao and A.~A. Malikopoulos, ``Enhanced mobility with connectivity and automation: A review of shared autonomous vehicle systems,'' \emph{IEEE Intelligent Transportation Systems Magazine}, vol.~14, no.~1, pp. 87--102, 2020.

\bibitem{phan2020intelligent}
D.~Phan, A.~Bab-Hadiashar, C.~Y. Lai, B.~Crawford, R.~Hoseinnezhad, R.~N. Jazar, and H.~Khayyam, ``Intelligent energy management system for conventional autonomous vehicles,'' \emph{Energy}, vol. 191, p. 116476, 2020.

\bibitem{van2018autonomous}
J.~Van~Brummelen, M.~O’Brien, D.~Gruyer, and H.~Najjaran, ``Autonomous vehicle perception: The technology of today and tomorrow,'' \emph{Transportation research part C: emerging technologies}, vol.~89, pp. 384--406, 2018.

\bibitem{ye2019automated}
Y.~Ye, X.~Zhang, and J.~Sun, ``Automated vehicle’s behavior decision making using deep reinforcement learning and high-fidelity simulation environment,'' \emph{Transportation Research Part C: Emerging Technologies}, vol. 107, pp. 155--170, 2019.

\bibitem{wang2023adaptive}
J.~Wang, X.~Yuan, Z.~Liu, W.~Tan, X.~Zhang, and Y.~Wang, ``Adaptive dynamic path planning method for autonomous vehicle under various road friction and speeds,'' \emph{IEEE Transactions on Intelligent Transportation Systems}, 2023.

\bibitem{hu2020fuzzy}
C.~Hu, Y.~Chen, and J.~Wang, ``Fuzzy observer-based transitional path-tracking control for autonomous vehicles,'' \emph{IEEE Transactions on Intelligent Transportation Systems}, vol.~22, no.~5, pp. 3078--3088, 2020.

\bibitem{hu2022decision}
Y.~Hu, L.~Yan, J.~Zhan, F.~Yan, Z.~Yin, F.~Peng, and Y.~Wu, ``Decision-making system based on finite state machine for low-speed autonomous vehicles in the park,'' in \emph{2022 IEEE International Conference on Real-time Computing and Robotics (RCAR)}.\hskip 1em plus 0.5em minus 0.4em\relax IEEE, 2022, pp. 721--726.

\bibitem{giunchiglia2019conditional}
E.~Giunchiglia, M.~Colledanchise, L.~Natale, and A.~Tacchella, ``Conditional behavior trees: Definition, executability, and applications,'' in \emph{2019 IEEE International Conference on Systems, Man and Cybernetics (SMC)}.\hskip 1em plus 0.5em minus 0.4em\relax IEEE, 2019, pp. 1899--1906.

\bibitem{zhu2018human}
M.~Zhu, X.~Wang, and Y.~Wang, ``Human-like autonomous car-following model with deep reinforcement learning,'' \emph{Transportation research part C: emerging technologies}, vol.~97, pp. 348--368, 2018.

\bibitem{balal2016binary}
E.~Balal, R.~L. Cheu, and T.~Sarkodie-Gyan, ``A binary decision model for discretionary lane changing move based on fuzzy inference system,'' \emph{Transportation Research Part C: Emerging Technologies}, vol.~67, pp. 47--61, 2016.

\bibitem{xiong2018decision}
G.~Xiong, Z.~Kang, H.~Li, W.~Song, Y.~Jin, and J.~Gong, ``Decision-making of lane change behavior based on rcs for automated vehicles in the real environment,'' in \emph{2018 IEEE Intelligent Vehicles Symposium (IV)}.\hskip 1em plus 0.5em minus 0.4em\relax IEEE, 2018, pp. 1400--1405.

\bibitem{li2021optimization}
B.~Li, T.~Acarman, Y.~Zhang, Y.~Ouyang, C.~Yaman, Q.~Kong, X.~Zhong, and X.~Peng, ``Optimization-based trajectory planning for autonomous parking with irregularly placed obstacles: A lightweight iterative framework,'' \emph{IEEE Transactions on Intelligent Transportation Systems}, vol.~23, no.~8, pp. 11\,970--11\,981, 2021.

\bibitem{han2017multi}
Z.~Han, D.~Wang, F.~Liu, and Z.~Zhao, ``Multi-agv path planning with double-path constraints by using an improved genetic algorithm,'' \emph{PloS one}, vol.~12, no.~7, p. e0181747, 2017.

\bibitem{claussmann2019review}
L.~Claussmann, M.~Revilloud, D.~Gruyer, and S.~Glaser, ``A review of motion planning for highway autonomous driving,'' \emph{IEEE Transactions on Intelligent Transportation Systems}, vol.~21, no.~5, pp. 1826--1848, 2019.

\bibitem{scheffe2022sequential}
P.~Scheffe, T.~M. Henneken, M.~Kloock, and B.~Alrifaee, ``Sequential convex programming methods for real-time optimal trajectory planning in autonomous vehicle racing,'' \emph{IEEE Transactions on Intelligent Vehicles}, vol.~8, no.~1, pp. 661--672, 2022.

\bibitem{fayazi2018mixed}
S.~A. Fayazi and A.~Vahidi, ``Mixed-integer linear programming for optimal scheduling of autonomous vehicle intersection crossing,'' \emph{IEEE Transactions on Intelligent Vehicles}, vol.~3, no.~3, pp. 287--299, 2018.

\bibitem{xu2019design}
S.~Xu and H.~Peng, ``Design, analysis, and experiments of preview path tracking control for autonomous vehicles,'' \emph{IEEE Transactions on Intelligent Transportation Systems}, vol.~21, no.~1, pp. 48--58, 2019.

\bibitem{hwang2017path}
C.-L. Hwang, C.-C. Yang, and J.~Y. Hung, ``Path tracking of an autonomous ground vehicle with different payloads by hierarchical improved fuzzy dynamic sliding-mode control,'' \emph{IEEE Transactions on Fuzzy Systems}, vol.~26, no.~2, pp. 899--914, 2017.

\bibitem{chen2019human}
Y.~Chen, C.~Hu, and J.~Wang, ``Human-centered trajectory tracking control for autonomous vehicles with driver cut-in behavior prediction,'' \emph{IEEE Transactions on Vehicular Technology}, vol.~68, no.~9, pp. 8461--8471, 2019.

\bibitem{wang2020autonomous}
Z.~Wang, J.~Zha, and J.~Wang, ``Autonomous vehicle trajectory following: A flatness model predictive control approach with hardware-in-the-loop verification,'' \emph{IEEE Transactions on Intelligent Transportation Systems}, vol.~22, no.~9, pp. 5613--5623, 2020.

\bibitem{liang2021distributed}
J.~Liang, Y.~Lu, G.~Yin, Z.~Fang, W.~Zhuang, Y.~Ren, L.~Xu, and Y.~Li, ``A distributed integrated control architecture of afs and dyc based on mas for distributed drive electric vehicles,'' \emph{IEEE transactions on vehicular technology}, vol.~70, no.~6, pp. 5565--5577, 2021.

\bibitem{ji2016path}
J.~Ji, A.~Khajepour, W.~W. Melek, and Y.~Huang, ``Path planning and tracking for vehicle collision avoidance based on model predictive control with multiconstraints,'' \emph{IEEE Transactions on Vehicular Technology}, vol.~66, no.~2, pp. 952--964, 2016.

\bibitem{paden2016survey}
B.~Paden, M.~{\v{C}}{\'a}p, S.~Z. Yong, D.~Yershov, and E.~Frazzoli, ``A survey of motion planning and control techniques for self-driving urban vehicles,'' \emph{IEEE Transactions on intelligent vehicles}, vol.~1, no.~1, pp. 33--55, 2016.

\bibitem{peng2022integrated}
J.~Peng, S.~Zhang, Y.~Zhou, and Z.~Li, ``An integrated model for autonomous speed and lane change decision-making based on deep reinforcement learning,'' \emph{IEEE Transactions on Intelligent Transportation Systems}, vol.~23, no.~11, pp. 21\,848--21\,860, 2022.

\bibitem{naveed2021trajectory}
K.~B. Naveed, Z.~Qiao, and J.~M. Dolan, ``Trajectory planning for autonomous vehicles using hierarchical reinforcement learning,'' in \emph{2021 IEEE International Intelligent Transportation Systems Conference (ITSC)}.\hskip 1em plus 0.5em minus 0.4em\relax IEEE, 2021, pp. 601--606.

\bibitem{li2022combining}
S.~Li, C.~Wei, and Y.~Wang, ``Combining decision making and trajectory planning for lane changing using deep reinforcement learning,'' \emph{IEEE Transactions on Intelligent Transportation Systems}, vol.~23, no.~9, pp. 16\,110--16\,136, 2022.

\bibitem{hawke2020urban}
J.~Hawke, R.~Shen, C.~Gurau, S.~Sharma, D.~Reda, N.~Nikolov, P.~Mazur, S.~Micklethwaite, N.~Griffiths, A.~Shah \emph{et~al.}, ``Urban driving with conditional imitation learning,'' in \emph{2020 IEEE International Conference on Robotics and Automation (ICRA)}.\hskip 1em plus 0.5em minus 0.4em\relax IEEE, 2020, pp. 251--257.

\bibitem{liu2019evaluating}
H.~Liu, R.~Yang, L.~Wang, and P.~Liu, ``Evaluating initial public acceptance of highly and fully autonomous vehicles,'' \emph{International Journal of Human--Computer Interaction}, vol.~35, no.~11, pp. 919--931, 2019.

\bibitem{ruijten2018enhancing}
P.~A. Ruijten, J.~M. Terken, and S.~N. Chandramouli, ``Enhancing trust in autonomous vehicles through intelligent user interfaces that mimic human behavior,'' \emph{Multimodal Technologies and Interaction}, vol.~2, no.~4, p.~62, 2018.

\bibitem{wei2021human}
C.~Wei, E.~Paschalidis, N.~Merat, A.~Solernou, F.~Hajiseyedjavadi, and R.~Romano, ``Human-like decision making and motion control for smooth and natural car following,'' \emph{IEEE Transactions on Intelligent Vehicles}, 2021.

\bibitem{emuna2020deep}
R.~Emuna, A.~Borowsky, and A.~Biess, ``Deep reinforcement learning for human-like driving policies in collision avoidance tasks of self-driving cars,'' \emph{arXiv preprint arXiv:2006.04218}, 2020.

\bibitem{xu2020learning}
D.~Xu, Z.~Ding, X.~He, H.~Zhao, M.~Moze, F.~Aioun, and F.~Guillemard, ``Learning from naturalistic driving data for human-like autonomous highway driving,'' \emph{IEEE Transactions on Intelligent Transportation Systems}, vol.~22, no.~12, pp. 7341--7354, 2020.

\bibitem{naumann2020analyzing}
M.~Naumann, L.~Sun, W.~Zhan, and M.~Tomizuka, ``Analyzing the suitability of cost functions for explaining and imitating human driving behavior based on inverse reinforcement learning,'' in \emph{2020 IEEE International Conference on Robotics and Automation (ICRA)}.\hskip 1em plus 0.5em minus 0.4em\relax IEEE, 2020, pp. 5481--5487.

\bibitem{kebria2019deep}
P.~M. Kebria, A.~Khosravi, S.~M. Salaken, and S.~Nahavandi, ``Deep imitation learning for autonomous vehicles based on convolutional neural networks,'' \emph{IEEE/CAA Journal of Automatica Sinica}, vol.~7, no.~1, pp. 82--95, 2019.

\bibitem{liang2022mas}
J.~Liang, Y.~Li, G.~Yin, L.~Xu, Y.~Lu, J.~Feng, T.~Shen, and G.~Cai, ``A mas-based hierarchical architecture for the cooperation control of connected and automated vehicles,'' \emph{IEEE Transactions on Vehicular Technology}, vol.~72, no.~2, pp. 1559--1573, 2022.

\bibitem{wang2023lane}
F.~Wang, T.~Shen, M.~Zhao, Y.~Ren, Y.~Lu, B.~Feng, and G.~Yin, ``Lane-change trajectory planning and control based on stability region for distributed drive electric vehicle,'' \emph{IEEE Transactions on Vehicular Technology}, 2023.

\bibitem{ziebart2008maximum}
B.~D. Ziebart, A.~L. Maas, J.~A. Bagnell, A.~K. Dey \emph{et~al.}, ``Maximum entropy inverse reinforcement learning.'' in \emph{Aaai}, vol.~8.\hskip 1em plus 0.5em minus 0.4em\relax Chicago, IL, USA, 2008, pp. 1433--1438.

\bibitem{zhang2019non}
K.~Zhang, J.~Wang, N.~Chen, and G.~Yin, ``A non-cooperative vehicle-to-vehicle trajectory-planning algorithm with consideration of driver’s characteristics,'' \emph{Proceedings of the Institution of Mechanical Engineers, Part D: Journal of automobile engineering}, vol. 233, no.~10, pp. 2405--2420, 2019.

\bibitem{liang2023robust}
J.~Liang, Y.~Lu, J.~Feng, G.~Yin, W.~Zhuang, J.~Wu, L.~Xu, and F.~Wang, ``Robust shared control system for aggressive driving based on cooperative modes identification,'' \emph{IEEE Transactions on Systems, Man, and Cybernetics: Systems}, 2023.

\bibitem{vicente2020linear}
B.~A.~H. Vicente, S.~S. James, and S.~R. Anderson, ``Linear system identification versus physical modeling of lateral--longitudinal vehicle dynamics,'' \emph{IEEE Transactions on Control Systems Technology}, vol.~29, no.~3, pp. 1380--1387, 2020.

\end{thebibliography}
% Generated by IEEEtran.bst, version: 1.14 (2015/08/26)

\vspace{11pt}
\begin{IEEEbiography}
[{\includegraphics[width=1in,height=1.25in,clip,keepaspectratio]{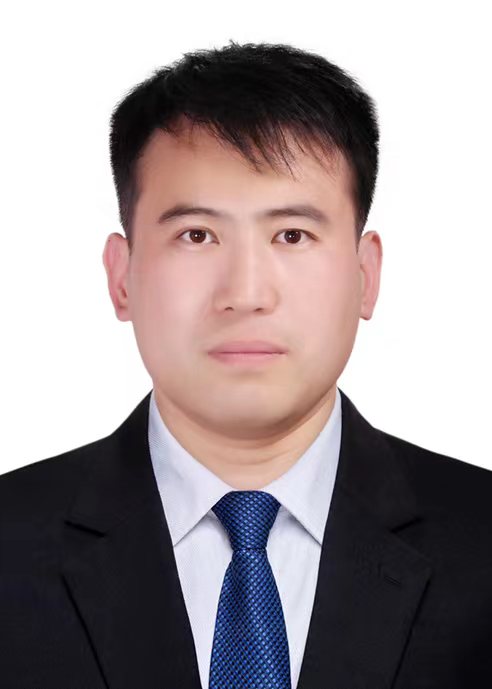}}]{Jinhao Liang}
received the B.S. degree from School of Mechanical Engineering, Nanjing University of Science and Technology, Nanjing, China, in 2017, and Ph.D. degree from School of Mechanical Engineering, Southeast University, Nanjing, China, in 2022. Now he is a Research Fellow with Department of Civil and Environmental Engineering, National University of Singapore. His research interests include vehicle dynamics and control, autonomous vehicles, and vehicle safety assistance system.
\end{IEEEbiography}
\vspace{11pt}

\begin{IEEEbiography}
[{\includegraphics[width=1in,height=1.25in,clip,keepaspectratio]{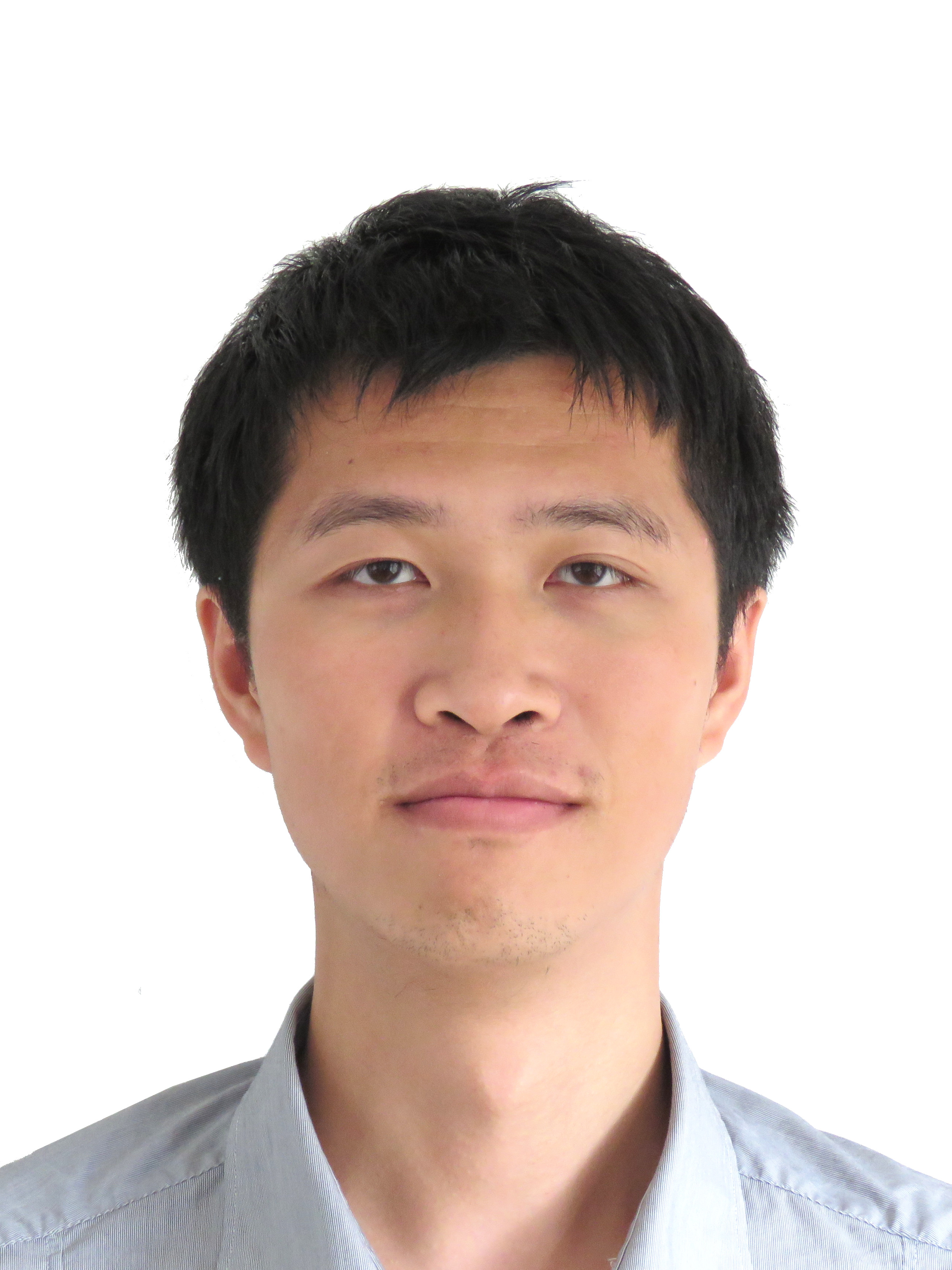}}]{Kaidi Yang} (Member, IEEE) is an Assistant Professor in the Department of Civil and Environmental Engineering at the National University of Singapore. Prior to this, he was a postdoctoral researcher with the Autonomous Systems Lab at Stanford University. He obtained a PhD degree from ETH Zurich and M.Sc. and B.Eng. degrees from Tsinghua University. His main research interest is the operation of future mobility systems enabled by connected and automated vehicles (CAVs) and shared mobility.
\end{IEEEbiography}
\vspace{11pt}
\begin{IEEEbiography}
[{\includegraphics[width=1in,height=1.25in,clip,keepaspectratio]{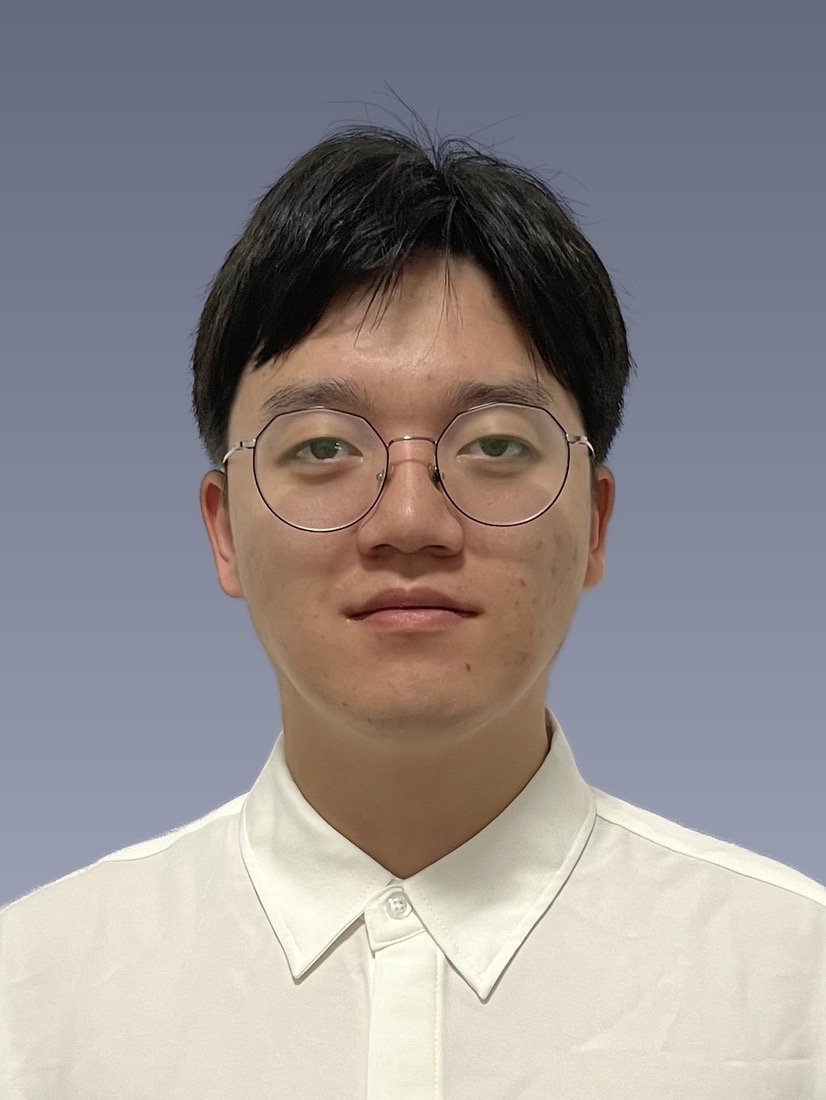}}]{Chaopeng Tan}
received his B.S. and Ph.D. degrees in traffic engineering from Tongji University, Shanghai, China in 2017 and 2022, respectively. He was also a visiting Ph.D. student at the University of Washington (Seattle) from 2019 to 2021. He is currently a Research Fellow with the Department of Civil and Environmental Engineering, National University of Singapore. His main research interests include intelligent transportation systems, traffic modeling and control with connected vehicles, and privacy-preserving traffic control.
\end{IEEEbiography}

\vspace{11pt}
\begin{IEEEbiography}
[{\includegraphics[width=1in,height=1.25in,clip,keepaspectratio]{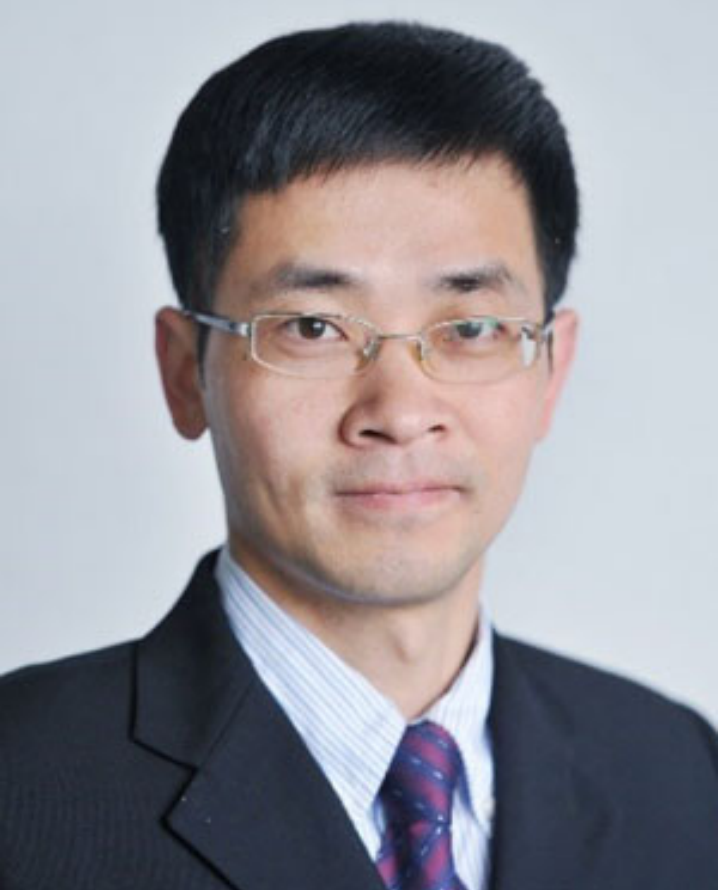}}]{Jinxiang Wang}
received the B.S. degree in mechanical engineering and automation, and the Ph.D. degree in vehicle engineering from Southeast University, Nanjing, China, in 2002 and 2010, respectively. From 2014 to 2015, he was a Visiting Research Scholar with the Department of Mechanical and Aerospace Engineering, The Ohio State University, Columbus, OH, USA. He is currently an Associate Professor with the School of Mechanical Engineering, Southeast University. His research interests include vehicle dynamics and control, assisted-driving system, and control on autonomous vehicles.
\end{IEEEbiography}

\vspace{11pt}
\begin{IEEEbiography}
[{\includegraphics[width=1in,height=1.25in,clip,keepaspectratio]{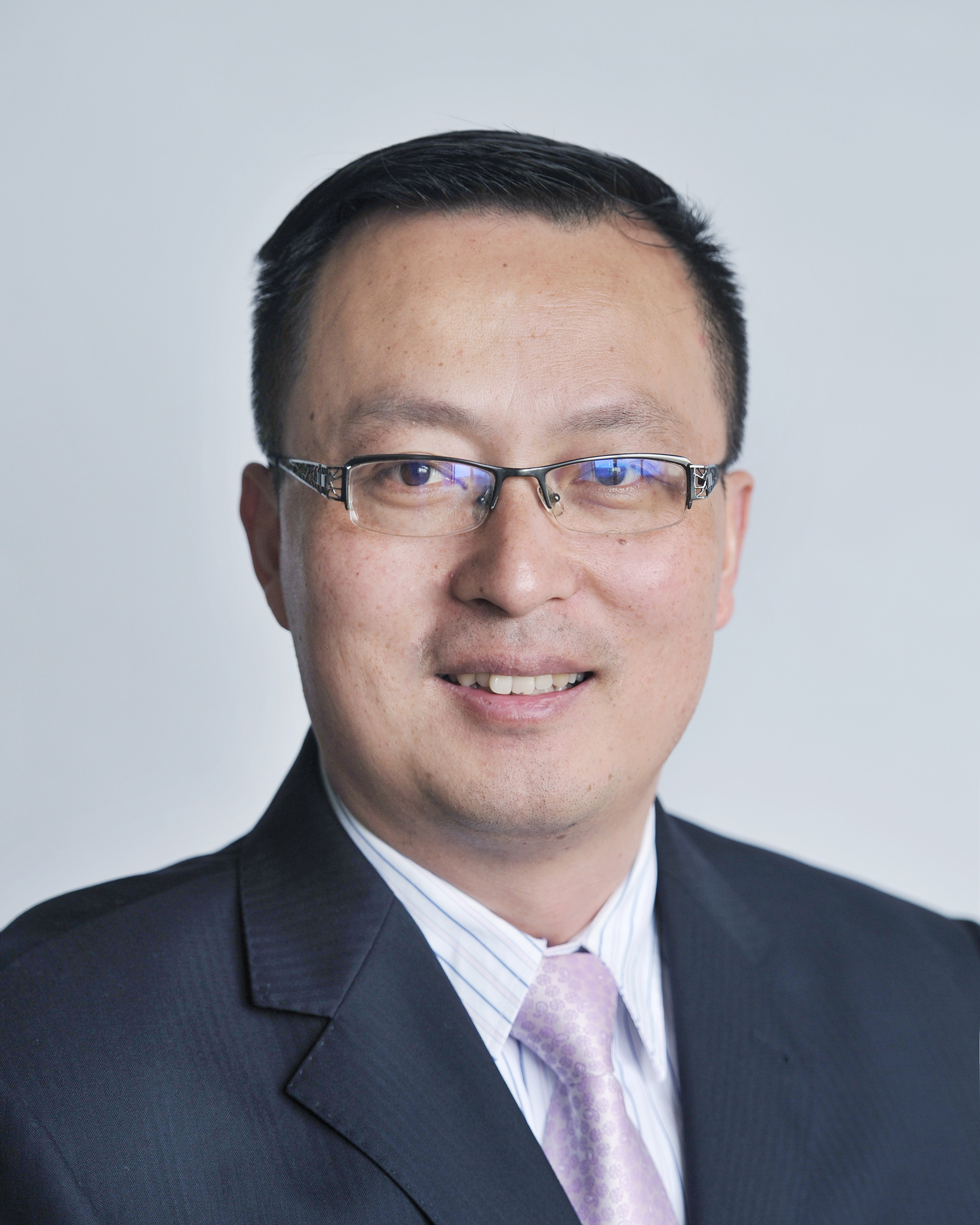}}]{Guodong Yin}
received the Ph.D. degree in mechanical engineering from Southeast University, Nanjing, China, in 2007. From August 2011 to August 2012, he was a Visiting Scholar with the Department of Mechanical and Aerospace Engineering, The Ohio State University, Columbus, OH, USA.
He is currently a Professor with the School of Mechanical Engineering, Southeast University. He was the recipient of the National Science Fund for Distinguished Young Scholars. His research interests include vehicle dynamics and control, automated vehicles, and connected vehicles.
\end{IEEEbiography}

\end{document}